\documentclass{article}

\usepackage{arxiv}

\usepackage[utf8]{inputenc} 
\usepackage[T1]{fontenc}    
\usepackage{hyperref}       
\usepackage{url}            
\usepackage{booktabs}       
\usepackage{amsfonts}       
\usepackage{nicefrac}       
\usepackage{microtype}      
\usepackage{lipsum}
\usepackage{graphicx}
\usepackage[dvipsnames]{xcolor}

\title{Time dependent lyotropic chromonic textures in PDMS-based microfluidic confinements}
\author{
Anshul Sharma, Irvine Lian Hao Ong and Anupam Sengupta \thanks{https://sites.google.com/site/anupamsengupta/} \\
 Physics of Living Matter, Department of Physics and Materials Science\\
  University of Luxembourg, \\
  162 A, Avenue de la Faencerie, L-1511, Luxembourg City, Luxembourg \\
  \texttt{anupam.sengupta@uni.lu} \\
}
\begin{document}
\maketitle

\begin{abstract}
Nematic and columnar phases of lyotropic chromonic liquid crystals (LCLCs) have been long studied for their fundamental and applied prospects in material science and medical diagnostics. LCLC phases represent different self-assembled states of disc-shaped molecules, held together by noncovalent interactions that lead to highly sensitive concentration and temperature dependent properties. Yet, microscale insights into confined LCLCs, specifically in the context of confinement geometry and surface properties, are lacking. Here, we report the emergence of time dependent textures in static disodium chromoglycate (DSCG) solutions, confined in PDMS-based microfluidic devices. We use a combination of soft lithography, surface characterization and polarized optical imaging to generate and analyze the confinement-induced LCLC textures, and demonstrate that over time, herringbone and spherulite textures emerge due to spontaneous nematic (N) to columnar M-phase transition, propagating from the LCLC-PDMS interface into the LCLC bulk. By varying the confinement geometry, anchoring conditions and the initial DSCG concentration, we can systematically tune the temporal dynamics of the N to M-phase transition and textural behaviour of the confined LCLC. Since static molecular states register the initial conditions for LC flows, the time dependent boundary and bulk conditions reported here suggest that the local surface-mediated dynamics could be central in understanding LCLC flows, and in turn, the associated transport properties of this versatile material.
\end{abstract}

\section{Introduction}
Lyotropic chromonic liquid crystals (LCLC) are a class of lyotropic liquid crystals (LCs), that are formed by anisotropic assemblies of water-soluble disc-shaped molecules that have an aromatic core surrounded by ionic groups. Most commonly used chromonic LCs are sunset yellow, an azo dye used as food additive and disodium chromoglycate (DSCG), a anti-asthmatic drug. Unlike lyotropic LCs, LCLCs do not form micelles, rather they stacks up as linear aggregates, held together by non-covalent interactions that lead to self-assembled nematic (N-phase) or columnar (M-phase) with hexagonal arrangement \cite{Lydon2010ChromonicReview,Lydon2011ChromonicPhases,Zhou2018RecentCrystals,Lubensky2017ConfinedMembranes}. In both N- phase and M-phase director is parallel to the columnar axis of the stacks, as was revealed by X-ray studies. The nematic phase consists of short columnar stacks of DSCG molecules and appears at room temperature for lower DSCG concentrations, whereas with increase in DSCG concentration the stacks assemble into a two dimensional hexagonal array of extended columns in the M-phase \cite{Hartshorne1973MesomorphismChromoglycate-water.,Cox1971SolidstateCromoglycate, Agra-Kooijman2014ColumnarCromoglycate}, archetypal herringbone and spherulite textures spanning a wide range of concentration and temperature conditions. The weak interaction forces underpinning the self-assembled phases render LCLCs highly responsive to external stimuli (temperature, concentration, pH, ionic content etc.) and geometric constraints \cite{Lydon2010ChromonicReview,Lydon2011ChromonicPhases,Zhou2018RecentCrystals,Lubensky2017ConfinedMembranes}, thus bestowing distinctive properties such as negative birefriengnece and large difference in elastic constants. Owing to biocompatibility and anisotropic properties, LCLCs have been explored in biological applications such as potential in drug delivery \cite{Simon2010NoncovalentIncompatibility}, optical biosensors \cite{Shiyanovskii2005LyotropicApplications}, as well in technological applications such as organic electronics and optical components \cite{Guo2011VerticallyPrecursors}. Lately, LCLCs have also been explored as host systems for active bacterial systems \cite{Mushenheim2014DynamicCrystals,Zhou2014LivingCrystals}. 

The challenges involved with alignment of LCLCs hinder its practical applications. Various techniques from traditional LCs alignment method such as rubbed polyimidie \cite{Collings2017AnchoringSurfaces}, external fields\cite{Lee1982LYOTROPICFIELDS.}, oblique evaporation \cite{Tone2012DynamicalCrystals}, self-assembled monolayers \cite{Simon2010ControllingNanotopography} etc. to sophisticated and complex methods such as graphene deposition \cite{Jeong2014HomeotropicInteractions}, sputter lithography \cite{Kim2016MacroscopicSubstrates}, photo-patterning \cite{Peng2017PatterningMetamasks} and plasmonic photopatterning of liquid crystal polymer network \cite{Dhakal2020Self-AssemblyBacteria} have been explored to achieve uniform alignment. Recently, study by Peng \textit{et al.} and Taras \textit{et al.} have shown that photopatterned orientation of LCLC can be used to control flow of active matter, that can be predesigned to streamline chaotic movements of swimming bacteria \cite{Peng2016CommandPatterns,Turiv2020PolarCrystal}. This works  demonstrates that precise control of LCLC alignment can be used to control orientation of active matter and has potential in microcargo delivery and soft microrobotics. However, to align DSCG perpendicular to the surface only few techniques such as silanes \cite{Nazarenko2010SurfaceCrystal}, hydrophobic polymers \cite{Tone2013}, graphene \cite{Jeong2014HomeotropicInteractions} and parylene \cite{Jeong2015ChiralCapillaries} have been successful. Lately, Guo and co-workers showed use of two-photon laser writing techniques to generate three dimensional topographical structures that can be used to uniformly and precisely control the alignment of LCLCs \cite{Guo2019PreciseTopography}.

In nematic phase, LCLCs are known to have an unusually low twist-elastic modulus\cite{Zhou2012ElasticityField,Nayani2015SpontaneousCylinders} and large saddle-splay modulus that breaks reflection symmetry and creates several new chiral director configurations when subjected to micro-confinement. Recently, few research groups have studied confinement induced reflection symmetry breaking and new chiral director configurations \cite{Tortora2011ChiralCrystals,Jeong2015ChiralCapillaries,Davidson2015ChiralElasticity,Nayani2015SpontaneousCylinders} of LCLCs. The chiral symmetry breaking for spherical droplets (tactoids) formed by both DSCG and SSY solutions of LCLC has been reported \cite{Jeong2014ChiralAnisotropy,Nayani2017UsingChromonics}. Jeong and co workers reported formation of hexagonally faceted droplet when system undergoes transition from nematic to columnar phase. When LCLCs are confined in glass capillaries with homeotropic boundary conditions, twisted escaped radial and twisted planar polar configurations have been reported. While a third chiral configuration, the escaped twist configuration, was observed in capillaries with degenerate planar boundary conditions \cite{Davidson2015ChiralElasticity,Nayani2015SpontaneousCylinders}. Dietrich \textit{et al.} have recently reported same chiral twisted escaped radial configuration in nematic phase of disc-like micelles of non-chromonic and achiral lyotropic LCs. A recent study showed that reflection symmetry breaking under confinement is not limited to LCLCs and suggested to be a general phenomenon exhibited by nematic  phase formed by supramolecular aggregates \cite{DIetrich2017ChiralConfinement}. Lately, Baza and co-workers have explored the dynamics of rheological properties of LCLCs under shear flow and have shown that DSCG exhibits tumbling characteristics similar to LC polymers \cite{Baza2020Shear-inducedCromoglycate}. 

Despite continued efforts to understand LCLC rheological properties, motivated by both fundamental and applied interests, we still lack a mechanistic insight into the LCLC hydrodynamics at microscales – a universal variable in medical, industrial and biological settings. With the advancement in soft lithography techniques which has enabled precise design and modulation of microfluidic geometry, pressure and boundary conditions, both biological and material technology have seen major advancements over the last decades \cite{Whitesides2006TheMicrofluidics}. Recent progress in liquid crystal microfluidics has demonstrated how hydrodynamics, in combination with surface anchoring and confinement can be harnessed to generate tunable flow and topological structures with potential for novel applications. The work of Sengupta \textit{et al.} \cite{Sengupta2011NematicEnvironmentb,Sengupta2013FlowMicro-pillar,Sengupta2015TopologicalProspects} and others \cite{Emersic2019SculptingLiquids} have shown that by precisely tuning the flow of nematic LCs, anchoring conditions at channel walls and confinement conditions in microchannels interesting phenomenon such as defect-mediated flow at low Reynolds's number \cite{Sengupta2012FunctionalizationFlows}, non-Poiseuille flow profile \cite{Sengupta2013LiquidShaping}, hydrodynamic cavitation \cite{Stieger2017HydrodynamicFluids} and crosstalk between topological field \cite{Giomi2017Cross-talkMicrofluidics} are observed. The work on nematic LCs in microconfinement has shown significant insight on the interplay of topological defects with tunable microflows of nematic LC and finds applications in micro cargo systems \cite{Sengupta2013TopologicalConcepts, Na2010ElectricallyArchitecture}, optofluidic system \cite{Cuennet2011OptofluidicMicroflows, Cuennet2013Optofluidic-tunableMicroflows,Wee2016TunableLens}, and fluidic resistance circuits \cite{Sengupta2013TuningMicrofluidics}. Based on these works, recently, flow-induced deformations have also been studied in chiral nematic LCs \cite{Wiese2016MicrofluidicCrystals, Guo2016CholestericStripes} and blue phases. Spontaneous tunable flow structures \cite{Sengupta2013LiquidShapingb,Sengupta2014LiquidMicroscales} were given a second look by Čopar and co-workers \cite{Copar2020MicrofluidicFlows}, revealing a hidden non-equilibrium chiral nematic state that could be stabilized by tuning microchannel dimensions and the driving pressure. 
\begin{figure}
\centering
\includegraphics[width=12cm]{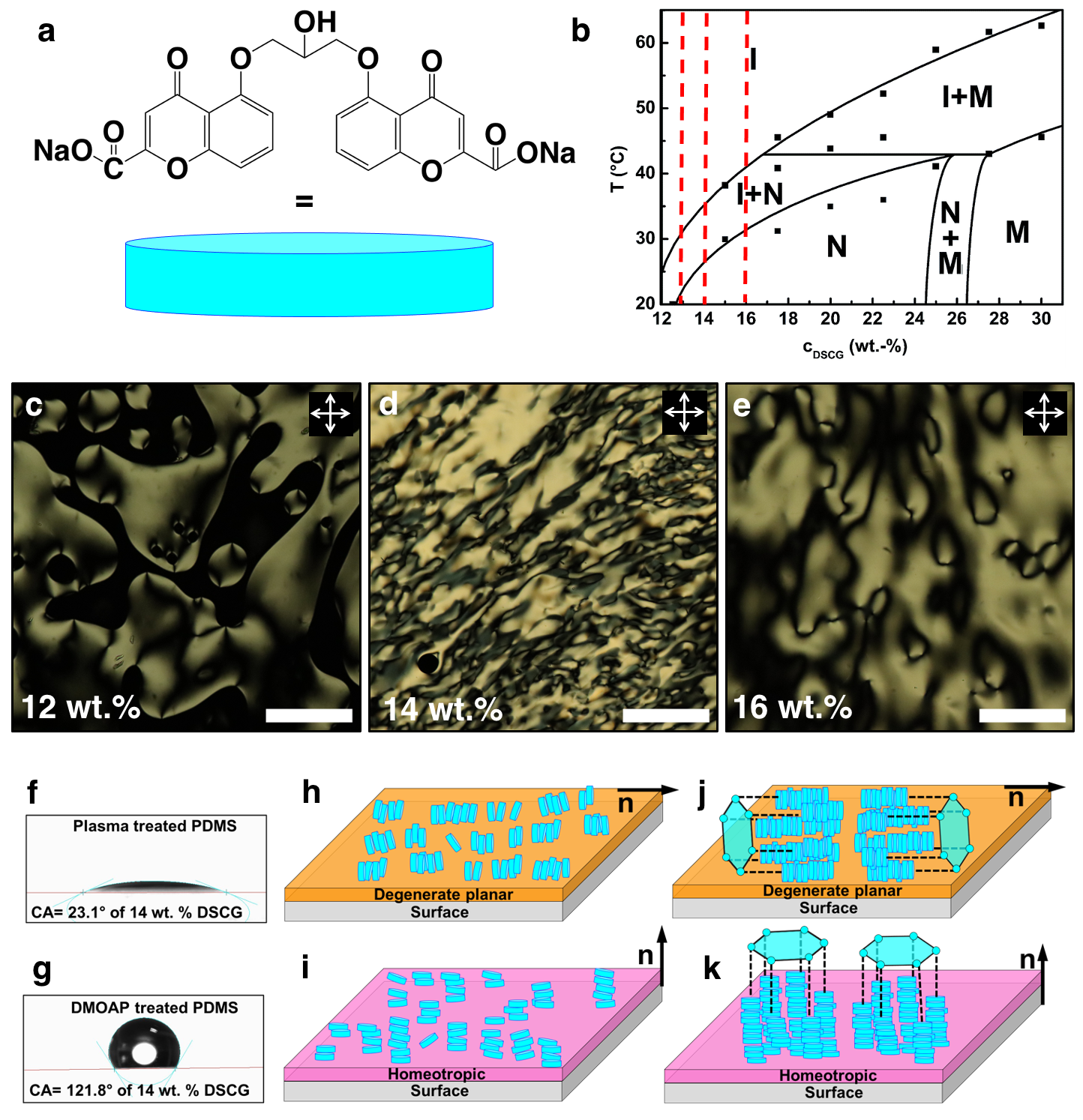}
\caption{\label{fig:Figure1-Schematic}\textbf{LCLC phases and surface-mediated alignment.} (\textbf{a}) Molecular structure of the DSCG molecule. (\textbf{b}), Phase diagram of DSCG in water (highlighted dotted red lines indicate three concentrations used in the current experiments), reproduced with permission from [\cite{Zimmermann2015Self-organizedDispersion}], published by The Royal Society of Chemistry. POM image of (\textbf{c}) isotropic and nematic droplets formed by 12~wt.~\% and (\textbf{d-e}) Schlieren texture of the nematic phase formed by 14~wt.~\% and 16~wt.~\% DSCG at 22~$^{\circ}$C, confined between untreated glass slide and cover slip (imaged between crossed polarizers, scale: 200~$\mu$m). (\textbf{f}) and (\textbf{g}) show the contact angles of the 14~wt.~\% DSCG with plasma- and DMOAP-treated PDMS surfaces, respectively. Schematic representation of the LCLCs: (\textbf{h}) and (\textbf{i}) for nematic and (\textbf{j}) and (\textbf{k}) columnar phase on degenerate planar alignment (\textit{edge-on} arranged molecules) and homeotropic alignment (\textit{face-on} arranged molecules), respectively.}
\end{figure}

The behaviour of LCLCs in strictly microfluidic environments still awaits exploration, with multiple key variables (that determine their interactions) yet to be uncovered and understood. As microfluidic environments offer distinct interplays between the surface, confinement and viscous parameters, one expects rich and unique interactions emerging at the LCLC-microfluidic interfaces. By substituting thermotropic nematic LCs with LCLCs within microfluidic confinements, the Authors have now opened a \textit{Pandora's box} wherein a wealth of novel phenomena are being observed which warrant fundamental investigations. Relying on the Authors' insights from previous experiments on thermotropic LCs in \textit{strictly} microfluidic confinements, here we continue with state-of-the-art soft lithography techniques to precisely tune the dimensions of the microchannels, and modify the surface anchoring conditions on the channel walls. It is known from study of thermotropic LCs in microfluidic environments that the initial conditions for flow-induced director response is defined by the static (no flow) conditions\cite{Sengupta2014LiquidMicroscales}. With a vision to comprehensively understand the flow-confinement-anchoring couplings in LCLCs confined within microchannels, here we analyze the fundamental \textit{no flow} state of LCLCs wihtin microfluidic enviroments. Specifically, we characterize how LCLC self-organizes within microscale confinements as a function of three fundamental variables: (i) channel dimensions, (ii) anchoring conditions, and (iii) LCLC concentration. In doing so, we have uncovered a fourth factor -- the experimental timescale -- which plays a critical role establishing the static director configurations within PDMS-glass microchannels. Our results reveal that the set of the four variable are fundamental in determining the outcome of flow experiments (flow results will be discussed elsewhere), and thus need to be taken into account when designing LCLC microfluidic experiments. In our first attempt to bridge the gap between LCLC dynamics and microfluidics, here we will focus on recent results on the temporal dynamics of LCLC textures formed by DSCG solutions confined in PDMS-glass microchannels, spanning different aspect ratios ($AR = width/height = w/h$) and anchoring conditions. The basic structure of DSCG, along with the stack configurations representing different phases are shown in Figure~\ref{fig:Figure1-Schematic}. 

\section{Materials and Methods}
Using a combination of precision microfluidics and polarization microscopy, we characterized the concentration-dependent static LCLC textures. We compared three concentrations of DSCG and characterized the LCLC optical textures as a function of anchoring conditions and channel aspect ratios, using polarizing optical microscopy and image analysis. Our results indicate that the emergence and orientation of the LCLC micro-domains are highly sensitive to confinement length scales and surface anchoring conditions. Additionally, a highlight of this work is the spontaneous nematic to columnar (M) phase transition in PDMS-glass microchannels, as a function of time (under steady state temperature conditions). The time of appearance of the M-phase shows a strong dependence on the channel geometry (width v/s depth) and surface anchoring (degenerate planar v/s homeotropic), indicating the fundamental role played by the properties of the confining surfaces (specifically, PDMS in this case) that need quantitative discussions detailed in this reference \cite{Sharma2020Surface-mediatedPrepration}. All our experiments were conducted in multiple replicates (minimum 3) to ensure reproducibility of the results. 

\subsection{Materials and solution preparation}
Disodium cromoglycate and dimethyloctadecyl[3-(trimethoxysilyl)propyl]ammonium chloride solution~(DMOAP) were purchased from Sigma-Aldrich. 4-cyano-4-pentyl-1,10-biphenyl (5CB), procured from Synthon Chemicals and used as received. All the solvents used are of GC purity~($\geq$~99.8~\%) sourced from Carl-Roth. Solutions with 12, 14 and 16~wt.~\% of DSCG were prepared in millipore water, vortexed for 30~minutes and stirred at room temperature for 12~hours to obtain a homogeneous solution. Two-component epoxy Araldit\textsuperscript{\textregistered} Rapid was purchased from Carl-Roth. Glass capillaries were acquired from CM Scientific Ltd. (Vitrocom Vitrotubes\textsuperscript{\textregistered}).

\subsection{Microfluidic confinement}
The experimental protocol used in the fabrication of PDMS microfluidic channels is based on the setup described previously \cite{Sengupta2012FunctionalizationFlows}. Briefly, microchannels were constructed using polydimethylsiloxane (PDMS, Sylgard 184, Dow Corning) reliefs and prepared by following the standard soft lithography techniques. The substrates were cleaned thoroughly with isopropanol and dried by placing the surfaces on a hot plate ($80^\circ$~C for 30~min). The molded PDMS reliefs were bonded to glass substrates after being exposed to oxygen plasma. We prepared linear channels with a rectangular cross-section. The channel depth ($d$) was varied between 10-40~$\mu$m, while the channel width ($w$) was varied between 100-345~$\mu$m. The distance between the inlet and the outlet port was set to 20~mm, defining the length(~$l$) of the channel. Different regions along the channel length were observed under microscope to estimate the uniformity of functionalization. At both ends of the microchannel, cylindrical holes with 1.5~mm diameter were punched through the PDMS to provide housing for the tubings. Two teflon (PTFE) tubes with 0.5~mm inner diameter and 1.58~mm outer diameter were inserted into each of housings and served as connectors for inlet-to-source and outlet-to-sink, respectively. For glass devices, hollow rectangle capillaries of dimensions (depth $\times$ width): 10 $\times$ 100~$\pm$~10~\%~$\mu$m, 40 $\times$ 400~$\pm$~10~\%~$\mu$m and 100 $\times$ 1000~$\pm$~10~\%~$\mu$m were used to study the texture development of LCLCs. Glass capillaries were either used as such or treated in order to obtain suitable surface anchoring and then filled with the liquid crystal. 

\subsection{Surface anchoring and boundary conditions}
Microfluidic devices were functionalized for investigating LCLCs textures within microchannels and glass capillaries possessing two distinct surface anchoring conditions: degenerate planar and homeotropic. Specific anchoring states within the microfluidic devices were achieved by a combination of distinct physical and chemical methods. We applied well-studied plasma exposure (inducing degenerate planar anchoring) and DMOAP treatment (homeotropic) for surface functionalization of the PDMS-glass microchannels and glass-only devices \cite{Sengupta2011NematicEnvironment,Sengupta2014LiquidMicroscales}. The anchoring conditions were validated using thermotropic nematic 5CB as our control sample, for which detailed studies exist \cite{Sengupta2013TopologicalEnvironment}. 

The degenerate planar surface anchoring is obtained after treatment of the PDMS surface by air plasma. Experiments under degenerate planar alignmnent conditions were conducted under freshly treated channels only. For homeotropic alignment the channels were first filled with 0.4~wt.\% aqueous solution of DMOAP and then dried in oven at $80^\circ$~C for 2~hours to render both the PDMS and glass surfaces hydrophobic, inducing perpendicular (homeotropic) boundary conditions. In order to achieve homeotropic anchoring of the director at the LC-glass interface, the glass capillaries were filled with a 0.4~wt.\% aqueous solution of DMOAP and the solvent was allowed to evaporate before the capillary was filled with the liquid crystal. The anchoring inside the capillaries were characterized through the shape of the meniscus of water and nematic 5CB, relative to the shape inside native capillaries, as shown in Figure~\ref{fig:Meniscus}.

\subsection{Filling the microfluidic devices}
Glass-PDMS and glass capillaries were filled above the isotropic-nematic transition temperature of LCLC solution for a given concentration of DSCG (12, 14 or 16 wt.~\%) at $45^\circ$~C either with a syringe and needle or capillary filling, respectively. Typically, a microchannel kept at $45^\circ$~C was filled with a syringe~(3ml) with bevelled tip needle~(23~gauge, 25~mm long, 0.6~mm outer diameter), to which a PTFE tube~(0.5~mm inner diameter and 1.58~mm outer diameter) was connected. The other end of the PTFE tube was inserted into microchannel. After filling, microchannel was sealed with epoxy glue, to avoid flow-induced effects (e.g., alignment) and, crucially, prevent water evaporation that could change the DSCG concentration. Thereafter, microchannels were slowly cooled down to $22^\circ$~C with a ramp rate of $1.5^\circ$~C per minute. The capillaries were filled with 14~wt.~\% DSCG solution by capillarity at $45^\circ$~C and then immediately sealed with epoxy to prevent water evaporation and to avoid flow-induced effects, and then allowed to slowly cool down to $22^\circ$~C with a ramp rate of $1.5^\circ$~C per minute.

\subsection{Optical microscopy and image acquisition}
The director orientation of LCLCs in the PDMS-microchannels and glass capillaries was determined using polarised light microscope (Eclipse LV100N-POL, Nikon), equipped with $\times$~5, $\times$~10, $\times$~20 and $\times$~40 objectives and with a Linkam~(PE120) heating/cooling stage. The samples were observed between crossed polarizers in a transmission mode using monochromatic light and several polarizer configurations. Additionally, the channels were oriented at $45^\circ$ with respect to the polarizers and with and without insertion of a 530~nm phase-retardation plate. The images~(6000 $\times$ 4000 resolution) and videos~(1920 $\times$ 1080 resolution) were acquired using using digital a Canon~EOS77D camera and videos were recorded in full HD colour at a frame rate of 25 frames per second. The image and video analysis was performed by using tools ImageJ\textsuperscript{\textregistered} and VideoMach\textsuperscript{\textregistered}, respectively.

\subsection{Scanning electron microscopy}
The morphological properties of pristine, oxygen plasma-treated and DMOAP-treated PDMS were analyzed by scanning electron microscopy (SEM). SEM imaging of  the PDMS microchannels was done using JEOL JSM-6010LA being operated in 12--15~kV range using an In-lens secondary electron detector. For SEM imaging samples are gold coated ($\approx$1~nm thickness) using a sputter coater (Quorum: Q150 Thin Film Sputter Coater) for 30-120~seconds.

\begin{figure}
\centering
\includegraphics[width=10cm]{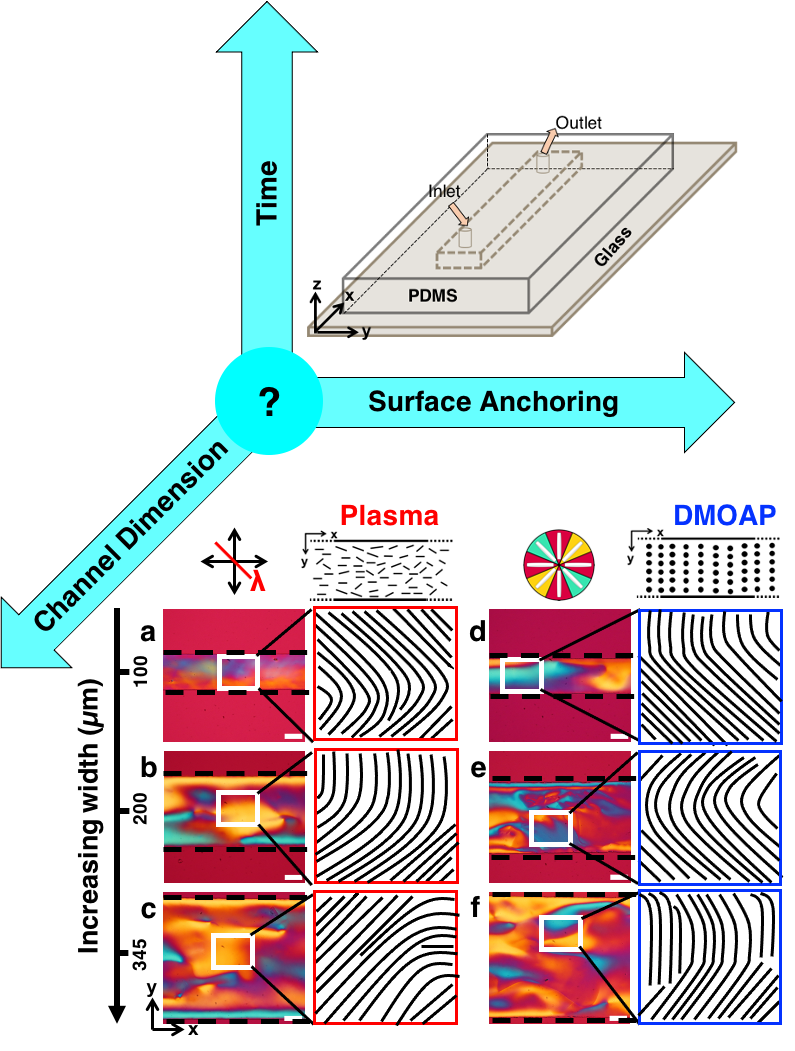}
\caption{\label{fig:Figure2-POM14DSCG-allPlasma-DMOAP} \textbf{Experimental parameters and microfluidic textures exhibited by 14~wt.~\% DSCG, shortly after the isotropic-nematic phase transition}, imaged between crossed polarizers with $\lambda$-retardation plate. (\textbf{a})-(\textbf{c}) Microchannels with degnerate planar anchoring with dimensions (from top): (100$\mu$m wide, 10$\mu$m deep), (200$\mu$m wide, 10$\mu$m deep) and (345$\mu$m wide, 10$\mu$m deep) respectively; and (\textbf{d})-(\textbf{f}) show corresponding  microchannels with homeotropic anchoring. The 2D  integrated director field of the LCLC in nematic phase (sketched alongside) indicates planar director configuration. Overall, no significant difference is observed between the two anchoring conditions during the short time span elapsed (order of few minutes) after the isotropic-nematic transition. Scale bar: 50 $\mu$m}
\end{figure}
\section{Results and Discussion}

The goal of this work is to study the equilibrium textures LCLC under strict microfluidic confinements possessing distinct surface anchoring conditions and channel dimensions. Figure~\ref{fig:Figure2-POM14DSCG-allPlasma-DMOAP} shows the experimental variables considered here: channel dimension, anchoring conditions and experimental timeline. For experiments conducted with PDMS-glass microchannels, all four walls possessed either a degenerate planar anchoring or homeotropic anchoring. From hereon, the results presented will focus on the concentration of DSCG to 14 wt.~\%, as our control case. Over the following sections, we present the dynamics of the emergence of the LCLC textures due to the DSCG solution. Based on the initial observations, a number of key questions--on the emergence of the dynamic and equilibrium LCLC textures--emerge, answering which could identify the specific roles of surface anchoring, confinement and LCLC concentrations. Below, we present a step by step, detailed report on the emergent textures as a function of the aforesaid factors.

\subsection{LCLC microfluidic textures due to degenerate planar anchoring}
As shown in Figures~\ref{fig:Figure2-POM14DSCG-allPlasma-DMOAP} (a-f), POM imaging was used to characterize the developing director orientation within the microchannels. To create the degenerate planar anchoring, we employed plasma treatment (discussed previously) as a way to modify the microchannel surfaces to align the nematic phase exhibited by 14 wt.~\% DSCG. The plasma-treated microchanel was the filled with DSCG, sealed and cooled to $22^\circ$~C with a ramp rate of $1.5^\circ$~C per minute. In order to perform preliminary optical characterization of the textures, we scanned the entire channel at a lower magnification (5$\times$), followed by images from selected areas (that best represented the overall director profile) were taken at a higher magnification (20$\times$), presented in Figure~\ref{fig:Figure2-POM14DSCG-allPlasma-DMOAP} (a-c). Corresponding textures in homeotropic channels are shown in Figure~\ref{fig:Figure2-POM14DSCG-allPlasma-DMOAP} (d-f), which interestingly, resemble the degenerate planar textures right after the isotropic to nematic phase transition. The anchoring condition was confirmed to be homeotropic by filling in 5CB and comparing the resulting texture with those in our previous works. 

\begin{figure}
\centering
\includegraphics[width=14cm]{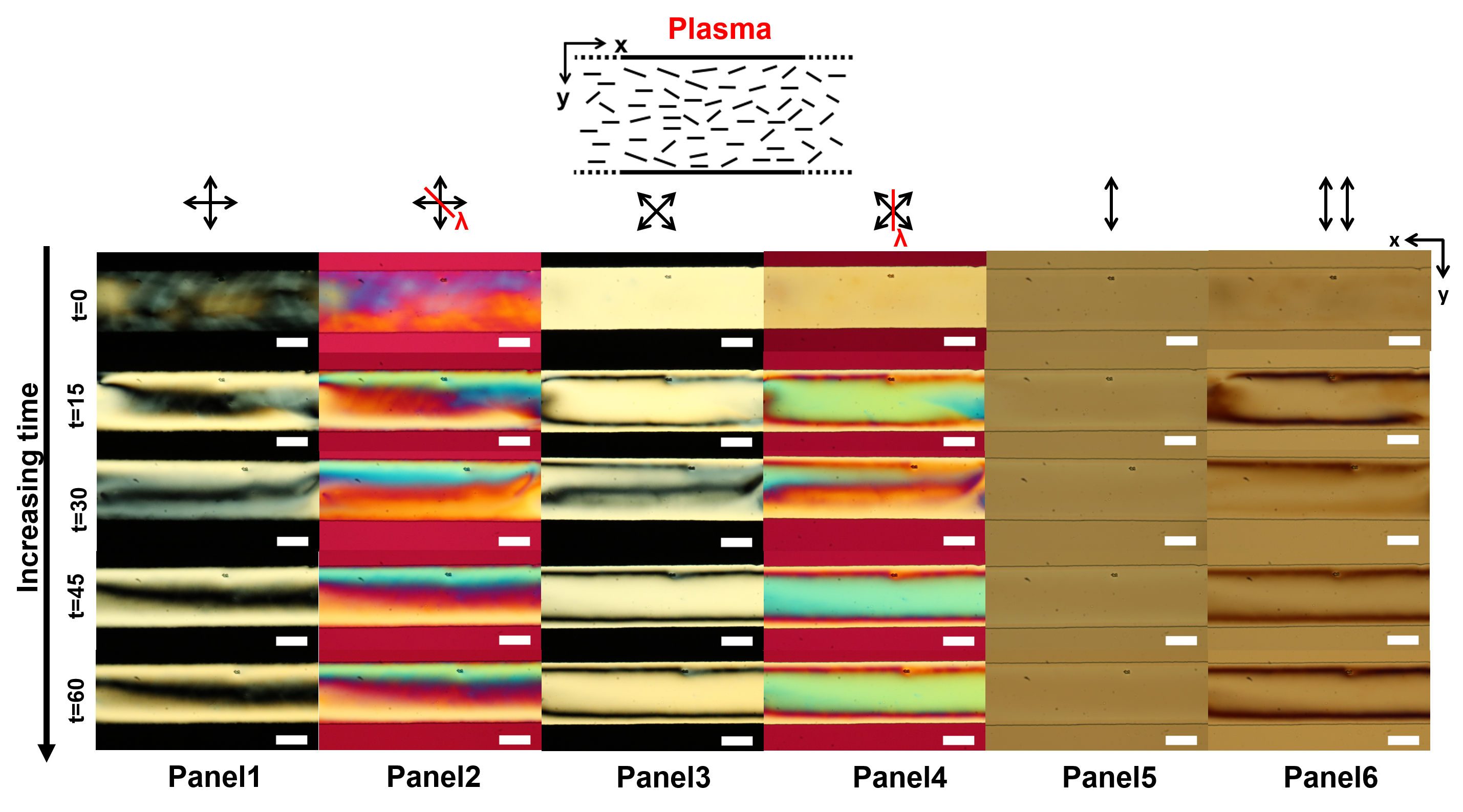}
\caption{\label{fig:plasma-100-10-1}\textbf{Nematic microfluidic textures due to degenerate planar anchoring}. Polarizing optical micrographs show the time evolution of the planar nematic textures in 14~wt.~\% DSCG after the isotropic-nematic transition, observed over a duration of 60 minutes, between: (\textbf{Panel 1}) crossed polarizers; (\textbf{Panel 2}) crossed polarizers and $\lambda$-retardation plate; (\textbf{Panel 3}) crossed polarizers with $45^\circ$ rotation; (\textbf{Panel 4}) crossed polarizers at $45^\circ$, with the $\lambda$-retardation plate; (\textbf{Panel 5}) analyzer only; and (\textbf{Panel 6}) parallel analyser and polarizer. Here, time (t) is in minutes after the first appearance of the nematic domains. The microchannels were 100$\mu$m wide and 10$\mu$m deep, scale bar: 50 $\mu$m.}
\end{figure}

Figure~\ref{fig:plasma-100-10-1} captures the time evolution of the DSCG textures within channels possessing degenerate planar anchoring. The dimension of the microchannel was kept fixed, with an $AR \approx10$ ($w$ = 100$\mu$m, $h$ = 10$\mu$m). Panels 1-6 (Figure~\ref{fig:plasma-100-10-1}), show POM micrographs obtained from a combination of different polarizer orientations, used to deduce the average molecular orientations corresponding to the planar textures. The birefringent domains appear shortly after the sample is cooled down below the isotropic-nematic phase transition temperature, and stabilize over a timescale of 60 minutes. This appears as dark and bright regions between crossed polarizers, which show the formation of randomly oriented planar DSCG domains. The textures were further characterized using a $\lambda$-retardation plate along with the crossed polarizers (panel 2). It may be worthwhile to note here that the N and M LCLC phases are known to posssess negative birefringence \cite{Nastishin2005OpticalParameter}, thus, in our POM micrographs with $\lambda$-retardation plates, the regions in yellow represent the nematic director orientation parallel to the slow axis of the retardation plate (please see Figure~\ref{fig:Figure2-POM14DSCG-allPlasma-DMOAP} for the reference colour wheel). The region, where the director is oriented perpendicular to the slow axis of the retardation plate appears blue \cite{Mirri2014StabilisationApplications}. The colour variation can be attributed to varying LC director orientations of microdomains formed by LCLC stacks in the nematic phase. In order to determine the stability of the planar textures, we checked the texture evolution every 15~minutes after the ramp has finished until 60~minutes. Our observations reveal that initially (right after transition, t=0 minute), the DSCG solution has degenerate planar orientations. Different parts of the microfludic channel were monitored to ascertain the homogeneity of the textures along the channel length. For the region of interest presented here, the planar nematic textures were found to be stable and homogeneous for about 60 minutes. 

\begin{figure}
\centering
\includegraphics[width=15cm]{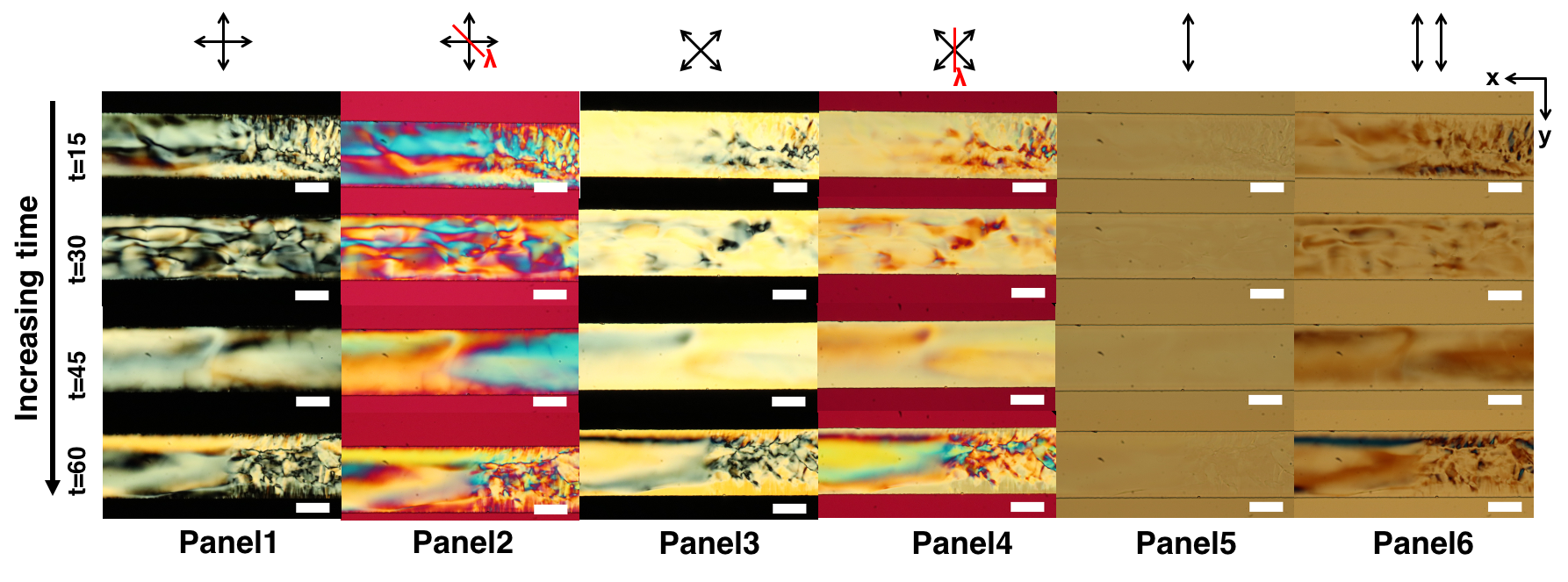}
\caption{\label{fig:plasma-100-10-2}\textbf{Appearance of the herringbone microfluidic texture due to the formation of M-phase in degenerate planar microchannels}. Panels 1-6 present the formation of the characteristic herringbone texture (in M-phase) in 14~wt.~\% DSCG, seen here on the right half of the micrographs. On the left half of the channel, DSCG can be seen to be still in the nematic phase. The micrographs were captured using different orientations of the polarizers relative to the channel, with or without the $\lambda$-retardation plate, as described in Figure~\ref{fig:plasma-100-10-1}. Here, time (t) is in minutes after the first appearance of the nematic domains. Microchannels were 100$\mu$m wide and 10$\mu$m deep, scale bar: 50 $\mu$m.}
\end{figure}

Interestingly, within similar timescales, over isolated regions within the microfluidic device, we observed the formation of $non$ planar static nematic textures (Figure~\ref{fig:plasma-100-10-2}). Starting from a degenerate alignment, such textures grew and propagated over time, ultimately stabilizing into \textit{herringbone} configuration, a classic LCLC M-phase texture (Figs.~\ref{fig:plasma-100-10-1},~\ref{fig:plasma-100-10-2} and~\ref{fig:M-phase-all} (Panel1). Tone \textit{et al.} have reported similar degenerate planar texture for a 13 wt.~\% DSCG solution in polylysine coated LC glass cell \cite{Tone2012DynamicalCrystals}. It was also observed that degenerate planar texture evolved to ribbon structure, texture corresponding to coexistence of M and isotropic phase. As there is no preferential alignment of the director along any direction, the DSCG stacks orient randomly forming a multi-domain structures of different orientations of the director. However, the timescale over which the planar textures transition from the nematic to herringbone (M-phase) depends on the channel aspect ratio and the DSCG concentration, as discussed in Section \ref{dimension}. 

\subsection{Nematic texture in homeotropic microchannels}
\begin{figure}
\centering
\includegraphics[width=14cm]{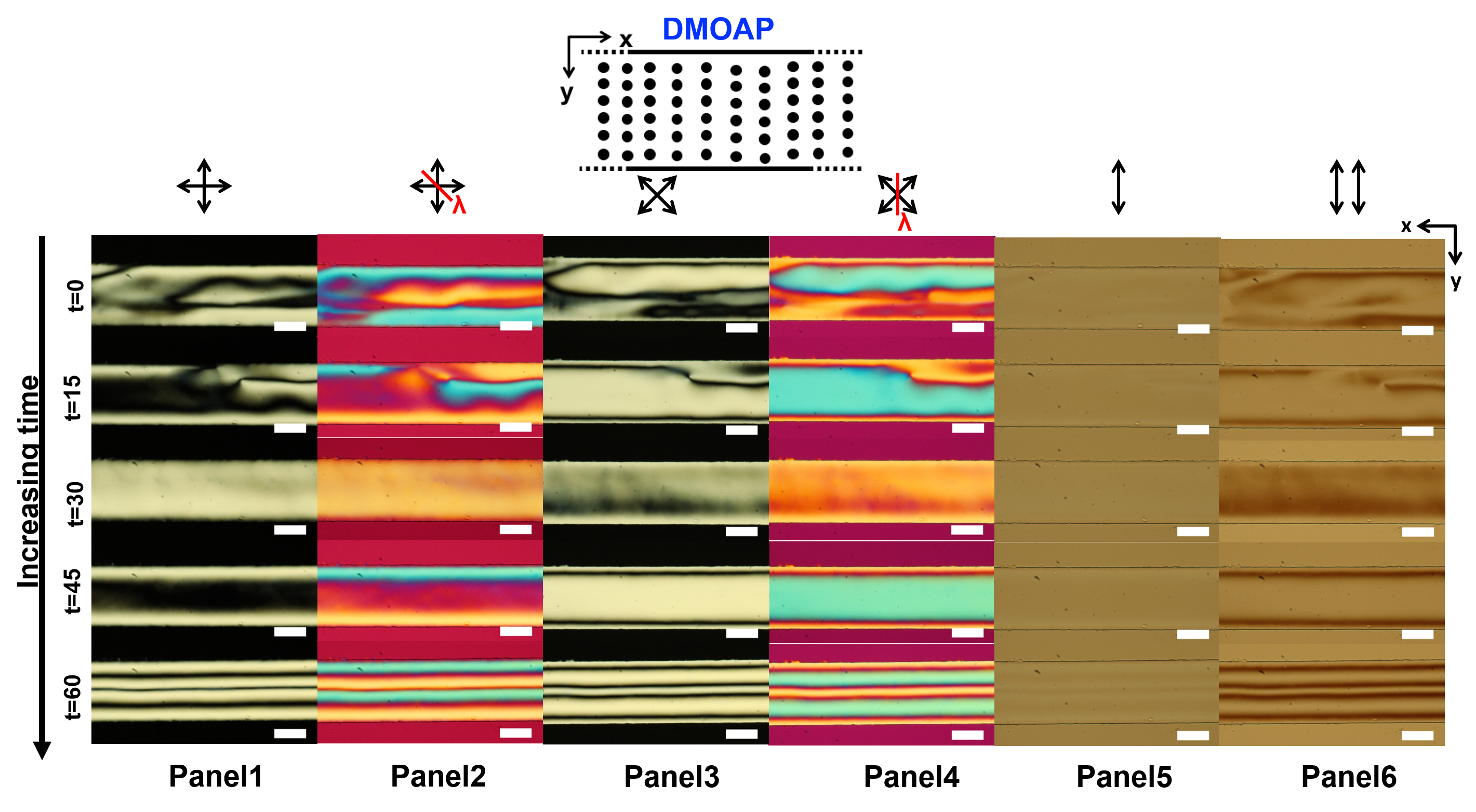}
\caption{\label{fig:DMOAP-100-10-1}\textbf{Nematic textures in microchannels with homeotropic surface anchoring}. Polarizing optical micrographs capture the time evolution (t=0 to 60 minutes) of LCLC texture within homeotropic PDMS-glass devices. The $pseudo$ $planar$ texture arises due to the spontaneous organization of the confined DSCG solution into planar or tilted orientation. (\textbf{Panels 1, 3}) are imaged between crossed polarizers with the channel oriented parallel and at $\pi/4$ to the polarizer, respectively;  (\textbf{Panels 2, 4}) micrographs imaged between crossed polarizers with an additional $\lambda$-plate; (\textbf{Panel 5}) with analyzer only; and (\textbf{Panel 6}) parallel analyser and polarizer, both perpendicular to the channel length. Here, time (t) is in minutes after the fist appearance of birefingence. Scale bar: 50 $\mu$m}
\end{figure}

We explored surface treatment with silanes as a way to modify the surface of PDMS-glass microchannels that support homeotropic anchoring. DMOAP-treated microchannels (prepared freshly) were filled with 14~wt.~\% DSCG solution in isotropic phase, sealed and then allowed to cool down to $22^\circ$~C, maintaining a steady  ramp rate of $1.5^\circ$~C per minute, as reported previously for the plasma-treated glass-PDMS microchannels. Figure~\ref{fig:DMOAP-100-10-1} summarizes the resulting LCLC textures produced by 14~wt.~\% DSCG solution within homeotropic PDMS-glass microchannels. The non-zero birefringence patterns suggest that the DSCG solution forms organize into unformly aligned planar or tilted domains within PDMS-glass microchannels, over the observed time window. 

By capturing the evolution of the textures over time, we observe that, initially, the LCLC nematic textures resemble the degenerate planar textures, even under homeotropic boundary conditions. We call this as the $pseudo$ $planar$ texture, arising due to planar or tilted organizatin of the DSCG nematic discotic stacks. This is in contrast to the microscale thermotropic nematic textures formed due to homeotropic boundary conditions: here the polarized light undergoes complete extinction upon passing through the sample (thus, appearing completely dark). 

\begin{figure}
\centering
\includegraphics[width=15cm]{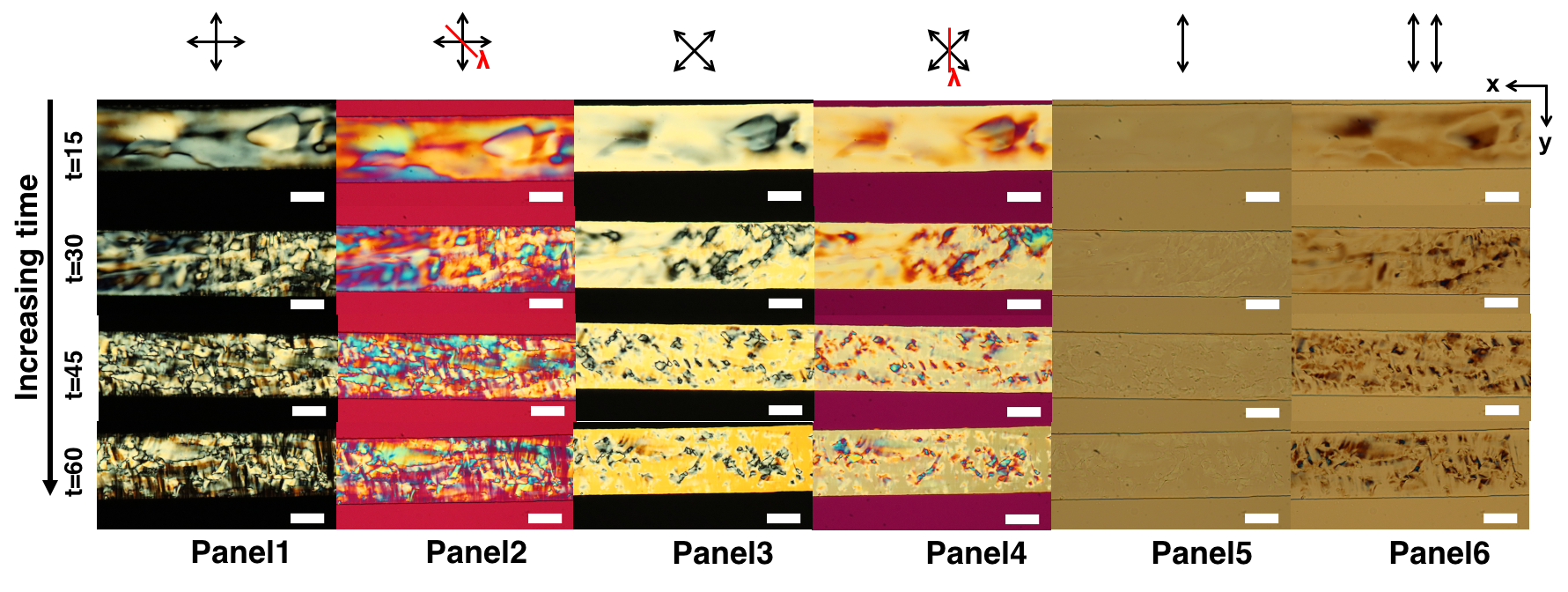}
\caption{\label{fig:DMOAP-100-10-2}\textbf{Transition of $pseudo$ $planar$ texture to spherulite texture} in 14~wt.~\% DSCG, confined within microchannels functionalized for homeotropic anchoring, as a function of time. Polarizing optical micrographs show the top view of the channel from t=15 to 60 minutes. The micrographs were captured using different orientations of the polarizers relative to the channel length, with or without the $\lambda$-retardation plate, as described in Figure~\ref{fig:DMOAP-100-10-1}. Scale bar: 50 $\mu$m.}
\end{figure}

As for the case of the degenerate planar anchoring (previous Section), for the homeotropic boudaries too, we observe isolated areas along the channel where the M-phase nucleates, and thereafter propagates through the channel, filling it up completely (Figure~\ref{fig:DMOAP-100-10-2}). However, unlike the degenerate planar case, the M-phase texture under the homeotropic boundary conditions manifests as the characteristic $spherulite$ texture. These contrasting outcomes--herringbone microfluidic texture in degenerate planar, and spherulite texture in homeotropic boundary conditions--suggest that the anchoring conditions play a key role in establishing the stable equilibrium static LCLC textures in PDMS-glass microfluidic confinements. Nazarenko and co-workers showed a weak homeotropic anchoring on hydrophobic silane-treated substrates \cite{Nazarenko2010SurfaceCrystal}, which eventually at a later time, transitioned from homeotropic to planar anchoring, while Tone \textit{et al.} reported that use of silane to get homeotropic alignmnet does not work on glass LC cells \cite{Tone2013}. From the texture evolution observed in Figure~\ref{fig:DMOAP-100-10-1} it can be concluded that for the tested silane ~(DMOAP)-treated PDMS-glass microchannel the nematic texture of DSCG does not show the typical homeotropic alignment and furthermore, the areas with homeotropic textures were found to be unstable. A time lapse sequence of the spontaneous transition from the $pseudo$ $planar$ to $spherulite$ texture is presented in Figure~\ref{fig:12DSCGplasma200-10-TL}.  

\subsection{\label{dimension}Role of the channel dimensions on the evolution of the LCLC textures}
The role of microfluidic dimensions was investigated by varying the channel aspect ratio, the ratio between the channel width $w$ and height $h$ ($AR = w/h$). We have studied the texture evolution by systematically varying both the channel width and height, through the following combinations of $w$ x $h$ (all dimensions in micrometers): (i) Variation of \underline{width} (in $\mu$m) - \underline{100} x 10, \underline{200} x 10 and \underline{345} x 10; and variation of \underline{height} (in $\mu$m) - 100 x \underline{10}, 100 x \underline{25} and 100 x \underline{40}. 

\subsubsection{Role of the channel width}
\begin{figure}
\centering
\includegraphics[width=12cm]{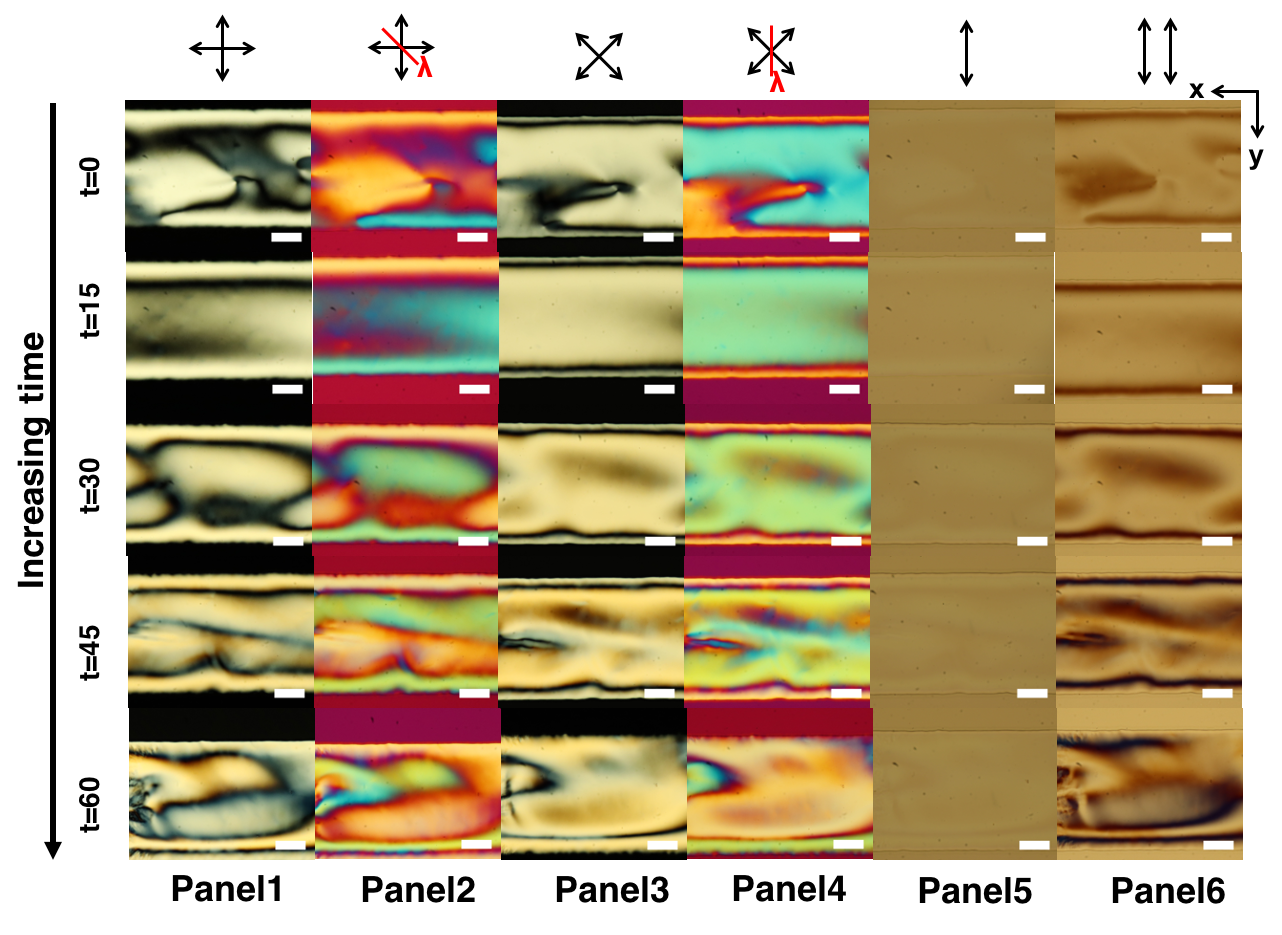}
\caption{\label{fig:plasma-200-10-1}\textbf{Effect of the channel width on the transition from schlieren to herringbone texture.} Evolution of the textures shown as a function of time, within 200 $\mu$m wide, 10 $\mu$m deep microchannel possessing degenerate planar anchoring, captured through polarized optical micrographs over t=0 to 60 minutes (the appearance of herringbone can be seen at t = 60 mins, on the left boundary of the image): (\textbf{Panel 1}) between crossed polarizers; (\textbf{Panel 2}) between crossed polarizers withe $\lambda$-plate inserted ; (\textbf{Panels 3 and 4}) similar to Panels 1 and 2, but with the crossed polarizers rotated by $45^\circ$; (\textbf{Panel 5}) micrograph with analyzer only; and (\textbf{Panel 6}) analyser parallel to the polarizer, but perpendicular to the channel length. Here, time (t) is in minutes after completion of the cooling ramp. Scale bar: 50 $\mu$m.}
\end{figure}

The role of channel width was investigated by studying two additional dimensions: 200$\mu$m and 345$\mu$m, while keeping the height fixed at 10$\mu$m. All other experimental conditions and parameters were kept similar to those used for the experiments reported in the previous Sections (with 100$\mu$m channel width). As described earlier, the stability of the degenerate planar anchoring was checked every 15~minutes after the cooling ramp concluded (this coincided with the first appearance of the birefringent domains).  For both the 200$\mu$m and 345$\mu$m wide microchannels, the observations lasted up to 60 minutes (or slightly longer, as the case was in a few instances), till the equilibrium LCLC static texture was observed, $i.e.$, no further change in the overall pattern and birefringence. The results are summarized in Figure~\ref{fig:plasma-200-10-1} using POM micrographs of the microchannels (200$\mu$m wide, 10$\mu$m deep) under different crossed polarizer orientations. 

\begin{figure}
\centering
\includegraphics[width=15cm]{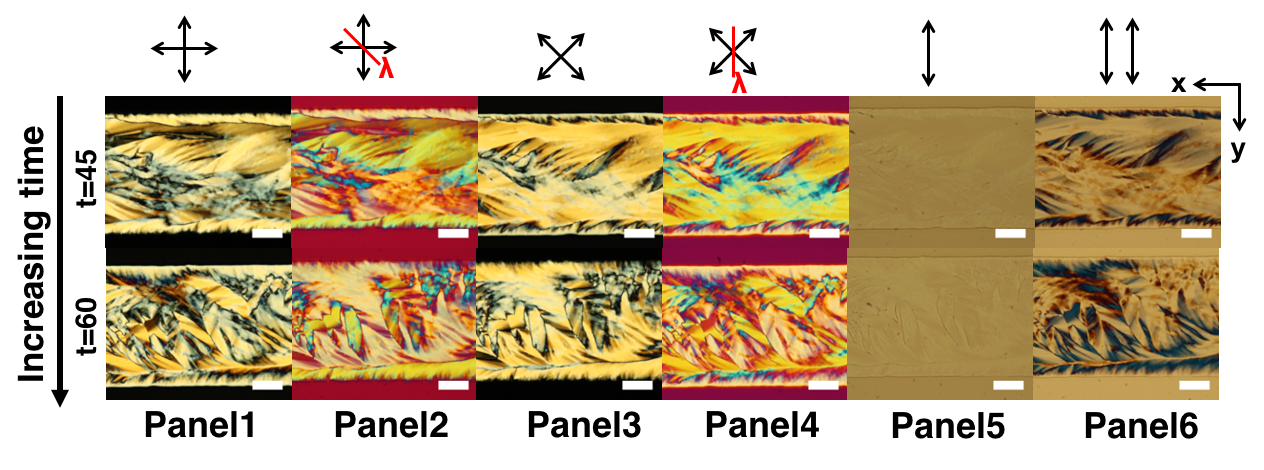}
\caption{\label{fig:plasma-200-10-2}\textbf{Characterization of the DSCG herringbone texture in (200$\mu$m wide, 10$\mu$m deep) microchannel with degenerate planar anchoring}. The fully developed M-phase texture was analyzed using POM under different polarization configurations, as described in previous images. The herringbone structure, globally, is oriented parallel to the microchannel length. Scale bar: 50 $\mu$m.}
\end{figure}

Initially, we observe the expected planar texture of DSCG due to the degenerate planar anchoring conditions on the channel walls.The planar microdomains, signifying different orientations, gave way to nucleation of the M-phase herringbone texture, wherein the discotic columns pack in more ordered two-dimensional lattice. Starting from a degenerate alignment (Figure~\ref{fig:plasma-200-10-1}), the herringbone texture proliferated along the channel length, and over a timescale of about 60 minutes, covered the entire channel (shown in Figure~\ref{fig:plasma-200-10-2}). It might be worth noting that the herringbone texture manifests a global orientation parallel to the length of the microchannel: this suggests that--under degenerate planar anchoring conditions--the microfluidic geometry underpins the emergent textural anisotropy of the confined DSCG (see Figure~\ref{fig:M-phase-all}). 

On increasing the width to 345 $\mu$m, as shown in Figures~\ref{fig:plasma345-10-1} and~\ref{fig:plasma345-10-2} we observed the textural transition from schlieren to herringbone texture consistently, however for wider channels the transition occurs over a shorter timescale. The relation between the transition time and channel aspect ratio will be reported elsewhere~\cite{Sharma2020Surface-mediatedPrepration}. In summary, the degenerate planar anchoring evolves into M-phase well-aligned herringbone texture over a time span of 60 minutes, such that wider the channels, longer is the transition time (positively correlated). 

Moving from degenerate planar anchoring to homeotropic anchoring conditions, initially the characteristic homeotropic textures were observed (complete extinction of the polarized light) within the 200 $\mu$m wide, 10$\mu$m deep DMOAP-treated microchannels. Over a timescale of 15 minutes, the homeotropic domains gradually transform into the $pseudo$ $planar$ texture, as shown in Figure~\ref{fig:DMOAP200-10-1}. Finally the planar texture transforms into the M-phase, over a timescale of $~$30 minutes. Apart from textures observed in area of microchannel as shown in Figure~\ref{fig:DMOAP200-10-1}, two other areas of this microchannel were observed: one where where M-phase (Figure~\ref{fig:DMOAP200-10-2}) and other with homeotropic texture (Figure~\ref{fig:DMOAP200-10-3}), developed at t=45 minutes. For areas where homeotropic texture developed, on observing between crossed polarizers and rotation of sample by $45^\circ$, intensity of transmission light remained low. This area with homeotropic alignmnet further propagated in the microchannel until t= 60 minutes but eventually changed to a degenerate planar texture. Another part of the microchannel where M phase appeared at t=30 minutes (Figure~\ref{fig:DMOAP-100-10-2}), after 45 minutes the observed area completely changed to M-phase. On changing the channel width further to 345 $\mu$m (Figures~\ref{fig:DMOAP345-10-1} and \ref{fig:DMOAP345-10-2}), qualitatively, we observed the same trend: regions of low intensity of transmitted light were first observed, which then transformed into the $pseudo$ $planar$ (within t= 15 minutes), which eventually changed to the M-phase at t= 45 minutes. It may be worthwhile to repeat here that, under homeotropic boundary conditions, the M-phase is distinct from that within degenerate planar microchannels: In hometropic channel, the spherulite texture--stretching transversely across the channel length--is observed, whereas within the degenerate planar channel, the herringbone structure organizes along the channel length. 

Taken together, these observations indicate that the channel geometry and surface anchoring play a key role in the development of the static LCLC textures. Overall, the change in the channel width impacts the nematic to M-phase transition time scale: an increase of 100~\% channel width (from 100 $\mu$m to 200 $\mu$m) lead to the time of appearance of M-phase increase from by 100~\% and 200~\% for plasma and DMOAP-treated microchannels, respectively. On further increasing the width by 245~\% (from 100 $\mu$m to 345 $\mu$m) the time of appearance of M-phase is increased by 200~\% for both plasma and DMOAP-treated microchannels. These results motivated us to investigate the effects of the change of the channel depth, while keeping the channel width constant at 100 $\mu$m and 200 $\mu$m.

\subsubsection{Role of the channel height}
The effect of the channel height was studied using microchannels with $h$ = 25 $\mu$m and 40 $\mu$m, while maintaining the channel width $w$ = 100 $\mu$m. The time evolution of the DSCG textures obtained for plasma-treated microchannels of dimension 25$\mu$m and 40$\mu$m via POM are shown in Figure~\ref{fig:plasma-100-25-1}-\ref{fig:plasma-100-25-2} and Figures~\ref{fig:plasma-100-40-1} and \ref{fig:plasma-100-40-2}, respectively. On increasing the depth by 150~\% ($i.e.$, from 10 $\mu$m to 25 $\mu$m deep), the M-phase appeared between 45-60 minutes, whereas on increasing the depth by 300~\% (100 $\mu$m wide, from 10 $\mu$m to 40 $\mu$m deep), the M-Phase appeared between 60-75 minutes. So, the appearance of the M-phase, and hence, the nematic to M-phase transition, is retarded due to the increase in the microchannel height. For the DMOAP-treated microchannels (hometropic anchoring), the time evolution of the textures are shown in Figures~\ref{fig:DMOAP-100-25-1} and~\ref{fig:DMOAP-100-40-1}, respectively. Here too, the transition time scale is stretched, relative to our control case: the M-phase appears at about 60 minutes within the 100$\mu$m wide, 25$\mu$m deep channel; while no phase change was captured within the first 60 minutes inside the 100 $\mu$m wide, 40 $\mu$m deep channel. We also studied time evolution of textures for channels with 15$\mu$m depth and 200$\mu$m width, both for plasma (Figure~\ref{fig:plasma-200-15-1}-\ref{fig:plasma-200-15-2}) and DMOAP (Figure~\ref{fig:DMOAP-200-15-1}) treatment. In general, we observed that upon increment of the channel depth from 10 to 40 $\mu$m, the DSCG texture changes from uniform to non-uniform appearance, for both the degenerate (Figure~\ref{fig:plasma-100-40-1}) and the homeotropic (Figure~\ref{fig:DMOAP-100-40-1}) boundary conditions. Similar observations on texture uniformity were made when channel depth was changed from 10$\mu$m to 15$\mu$m for 200$\mu$m wide channel, as shown in Figure~\ref{fig:plasma-200-15-1} and Figure~\ref{fig:DMOAP-200-15-1}.

\begin{figure}
\centering
\includegraphics[width=15cm]{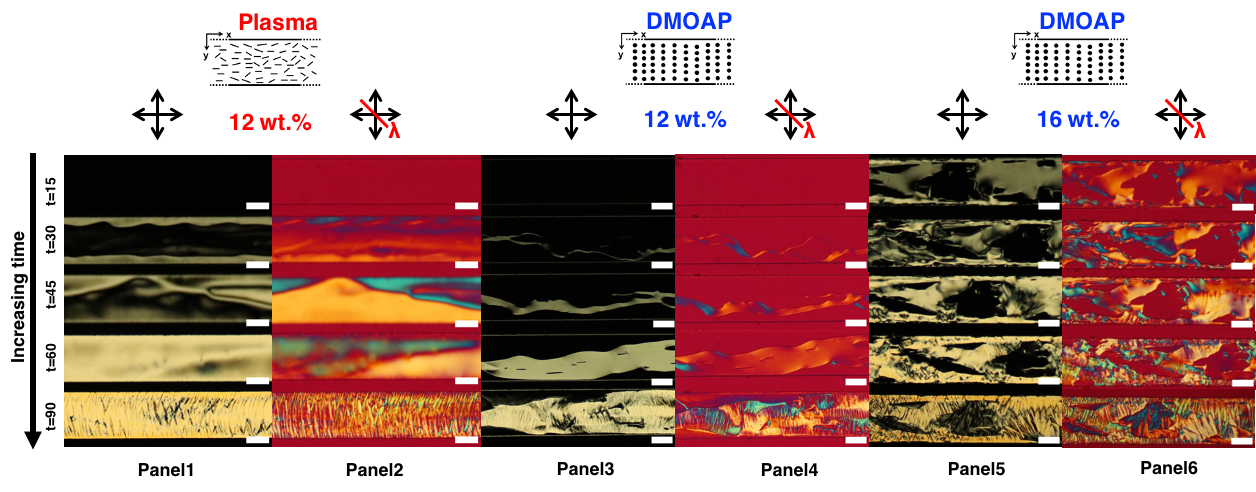}
\caption{\label{fig:POM12-16plasmaDMOAP}\textbf{DSCG textures at 12 wt.~\% and 16 wt.~\%, within 200$\mu$m wide, 10$\mu$m deep microchannel}. (\textbf{Panel 1-2}) and (\textbf{Panel 3-4}) Textures exhibited by nematic phase formed by 12~wt.~\% DSCG, under degenerate planar and homeotropic anchoring conditions respectively. (\textbf{Panel 5-6}) Textures exhibited by nematic phase formed by 16~wt.~\% DSCG, under homeotropic anchoring, observed through crossed polarizers and $\lambda$-retardation plate. Here, time (t) is in minutes after filling and sealing of glass-PDMS microfluidic device. Scale bar: 200 $\mu$m.}
\end{figure}

Based on the above results, we conclude that the stability of the nematic and M-phase textures, and the transitions therein, depend closely on the channel dimensions and the surface anchoring conditions. Furthermore, the height of the microchannel regulates the uniformity of the emergent textures: lower the height ($i.e.$, stronger confinement) leads to uniform texture, whereas in deeper channels the textures become increasingly non-uniform. Additionally, we also used two other DSCG concentrations for our experiments: 12 wt.~\% (closer to the isotropic phase at the operating temperature) and 16 wt.~\% (closer to M-phase) and studied the POM textures 200 $\mu$m wide and 10 $\mu$m microchannels. As shown in Figure~\ref{fig:POM12-16plasmaDMOAP} (panels 1-4), initially no microfluidic texture was detected at 12 wt.~\% DSCG, under both degenerate planar and homeotropic boundary conditions. Then, between 45-60 minutes, planar nematic textures emerged, which at about 90~minutes changed to the M-phase texture. At 16 wt.~\%, within DMOAP-treated microchannel (panels 5-6), a patchy homeotropic domain was observed (with regions of total light extinction), while rest of the regions were birefringent, which ultimately transformed to M-phase at around 60 minutes.

\subsection{Effect of surface treatment on LCLC microfluidic textures }
\begin{figure}
\centering
\includegraphics[width=14cm]{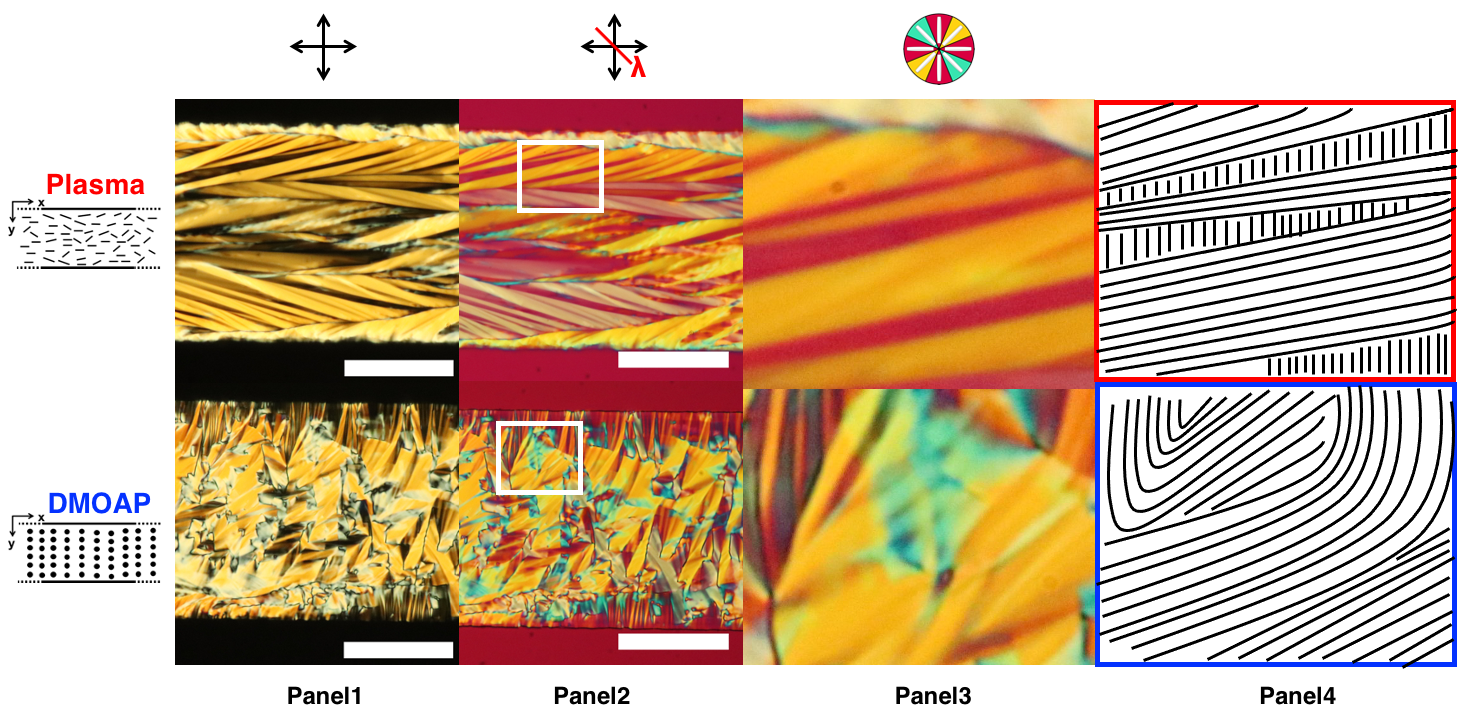}
\caption{\label{fig:Figure6-Mphase}\textbf{Herringbone and spherulite microfluidic textures in 14~wt.~\% DSCG.} (\textbf{Panel 1}) with crossed polarizers and (\textbf{Panel 2}) crossed polarizers with a $\lambda$-retardation plate show the microscale textures within the 200$\mu$m wide, 10$\mu$m deep, under degenerate planar and homeotropic anchoring respectively. (\textbf{Panel 3}) presents a zoomed in region, and (\textbf{Panel 4}) shows the schematic representation of the M-phase director field in 2D. Scale bar: 100 $\mu$m}
\end{figure}

Figure~\ref{fig:Figure6-Mphase} and Figure~\ref{fig:M-phase-all} shows the M-phase texture developed in PDMS-glass microchannels, under both plasma (degenerate planar) and DMOAP (homeotropic) surface treatments. For the microchannels with plasma treatment, the M-phase yields the herringbone texture, with an overall structural orientation along the channel lenght. For the DMOAP-treated microchannel, the M-phase elicits the spherulite texture, which aligns perpendicular to channel walls, $i.e.$, along the transverse direction (Figure~\ref{fig:Figure6-Mphase} and Figure~\ref{fig:M-phase-all}). As indicated by contact angle measurement of 14~wt.~\% DSCG solution (Figure~\ref{fig:Figure1-Schematic}, panels f and g), plasma exposure renders both PDMS and glass surfaces hyrophilic, whereas the silane, DMOAP, makes the surface hydrophobic. Consequently, nucleation, growth and propagation of the textures will be locally influenced by the immediate boundary conditions. In the planar case, the discotic stacks in the nematic or M-phase are expected to align along the planar domains imposed by the plasma treatment. This, with the additional constraint from the long axis of the microchannel, guides the herringbone structure along the channel length. For the homeotropic case, the spherulite texture emerges due to normal alignment of the discotics on the channel surfaces, and are thus constrained to grow from ceiling to the floor, and from wall-to-wall of the microchannel. Thus, we do not see any structural bias along the channel length. 

In order to gain a mechanistic insight, we looked into impact of oxygen plasma and DMOAP treatments on the PDMS surface morphology. Figure~\ref{fig:SEM} shows topography of the pristine, oxygen plasma-treated and DMOAP-treated PDMS. It has been known that plasma treatment of PDMS surface causes nano-structuring of PDMS and increase water contact angles depending on duration and intensity of plasma treatment used \cite{Bodas2007HydrophilizationInvestigation,Mata2005CharacterizationMicro/Nanosystems,Bhattacharya2005StudiesStrength,Hillborg2000CrosslinkedTechniques}. Bodas \textit{et al.} showed the notable cracks (through SEM) to the PDMS upon 150 W, 15 min, plasma treatment \cite{Bodas2007HydrophilizationInvestigation}. On the other hand, Hillborg \textit{et al.} employed 40 W, 3 min, plasma treatment and observed insignificant surface morphological changes (through SEM and AFM) \cite{Hillborg2000CrosslinkedTechniques}. From our SEM images, no significant change is visible between the untreated and plasma-treated PDMS; however the DMOAP-treated PDMS shows a non-uniform patterned surface. On looking mainly from the surface hydrophilicity/hydrophobicity angle (which is a robust and reliable outcome of oxygen plasma treatments on PDMS), Ruben \textit{et al.} had shown the enhanced capillary flow within PDMS microchannels that are hydrophilic \cite{Ruben2017OxygenMicrochannels}. Looking from a membrane point of view, hydrophobic species can permeate readily through PDMS, ranging from pharmaceutical compounds to organic dyes \cite{Toepke2006PDMSApplications,Waters2017EffectCompounds} but it is highly size dependant. Since, DSCG is a bulkier molecule with much higher solubility value (195.3 mg/ml \cite{Yoshimi1992ImportanceBioavailability}) compared to small molecules like Quinine or Nile red, we can rule out the possibility of DSCG permeation in PDMS. Nonetheless, in relation to DSCG, the absence of birefringent patterns in bulk PDMS is an additional support that DSCG does not permeate into the PDMS bulk. It can be concluded that a more hydrophilic PDMS surface would enhance the permeation of water through PDMS.

It is evident from the time-dependent optical observations that the nematic phase of 14~wt.~\% DSCG, within both plasma and DMOAP-treated microchannels, exhibited a phase transition from the nematic to the more ordered columnar M-phase. It is known that M-phase occurs at higher concentration of DSCG and for an N-M phase transition to occur from a nematic phase, there has to be change in concentration of DSCG (unless the temperature of the environment changes, which in our experiments, was kept fixed). The change in DSCG concentration can happen either on changing temperature or loss of water. As all the measurements were done at a constant temperature, the change in concentration of initial DSCG solution has to come from loss of water. In order to ensure that the change in textures of nematic phase to M-phase is a result of water loss due to the PDMS-glass microchannels, we compared the temporal stability of the DSCG textures (for 14~wt.~\% DSCG solution) within glass capillaries, for both native and DMOAP-treated samples. As shown in Figures~\ref{fig:POM14DSCGcapillarynotreatment} and~\ref{fig:POM14DSCGcapillaryDMOAP}, for untreated and DMOAP-treated glass capillaries respectively, the textures for planar and homeotropic alignment were observed. The time dependant texture analysis showed no N-M phase transition, regardless of the surface treatment. We also studied nematic phase of 14~wt.~\% DSCG solution by keeping one end of glass capillary open to allow water evaporation. Although the time evolution of texture showed a texture with birefriengence change, no herringbone M-phase texture was observed. This is shown in Figure~\ref{fig:Openglasscapillary}. The above combination of control tests confirms that the M-phase development within the PDMS-glass microchannels is mediated by the surface attributes of PDMS.

We studied 12 wt.~\% DSCG solution in a plasma-treated PDMS-glass microchannel Figure~\ref{fig:12DSCGplasma200-10-TL}. As this concentration is close to the isotropic-nematic phase traition temperatire (see, phase diagram in Figure~\ref{fig:Figure1-Schematic}), one should expect that the system has to first reach a concentration of DSCG that shows a stable nematic phase, and then the N-M phase transition follows. We filled a plasma-treated PDMS-glass microchannel with 12 wt.~\% solution and sealed it, and as discussed previously, carried out POM time lapse imaging (imaged every 30 seconds). As hypothesized, initially the system was observed to be in the isotropic phase, with the nematic phase emerging after about 25 minutes. Thereafter, the system stays in nematic phase for about 255 minutes, before the M-phase finally appears. This confirmatory test consolidates our conclusions that the surface properties of PDMS induces systematic changes in the initial concentration of DSCG, which ultimately lead to the formation of the spontaneous, yet tunable, LCLC phase transitions. A detailed quantitative analysis of the surface-mediated phase transition will be discussed elsewhere~\cite{Sharma2020Surface-mediatedPrepration}.

\section{Conclusions}
We exploit a simple and flexible soft lithography technique to create PDMS-glass microfluidic devices for studying the nucleation, growth and stability of LCLC textures. The temporal study carried out here reveals spontaneous (and unexpected) emergence of M-phase textures, with DSCG solutions which are originally in nematic phase. The comprehensive polarization optical study has demonstrated that static LCLC microfluidic textures are far from trivial, and are finely tuned by three key parameters:\\ 
\emph{(i) microchannel dimensions} (confinement effect),\\
\emph{(ii) surface anchoring} (anchoring effect), and crucially \\
\emph{(iii) experimental timescale} (spontaneous surface-mediated effects)\\
The three fundamental variables determine the local and temporal $no$-$flow$ LCLC microfluidic textures, giving us a set of key experimental parameters that could be potentially employed to register the initial conditions for microfluidic $flow$ experiments. 

Table~\ref{N-M phase transition time} summarizes the textural manifestations of DSCG solutions in PDMS-glass microchannels, and the timescales over which various surface-induced phase transitions occur. Multiple experimental replicates confirm the reproducibility of the values included in the Table~\ref{N-M phase transition time}, for a range of aspect ratios ($w/h$) of channels possessing either degenerate planar or hometropic boundary conditions. Although plasma treatment of microchannels successfully imposed degenerate planar conditions (for both thermotropic and  LCLC nematic phase), the well-known DMOAP treatment failed to generate a true homeotropic alignment of DSCG at 12 and 14 wt.~\% concentrations. However, at 16 wt.~\%, we observed larger homeotropic patches (leading to higher cumulative area showing homeotropic texture), with longer temporal stability.
\begin{table}
\caption{\label{N-M phase transition time} Impact of change in \textbf{\textit{channel dimension}} (variation in depth shown as underlined), \textbf{\textit{surface anchoring}} (\textcolor{orange} {degenerate planar} vs. \textcolor{magenta} {homeotropic}) and \textbf{\textit{DSCG concentration}} (12 , \textcolor{cyan} {14} and \textcolor{green} {16} wt.\%) on N-M phase transition time.}
\centering
\begin{tabular}{cccccc}
\toprule
\textbf{Channel dimension} & \textbf{Anchoring Condition} & \textbf{DSCG concentration} & \textbf{N to M phase transition}\\
\textbf{w $\times$ \underline{h} ($\mu$m)} & \textbf{} & \textbf{(\%)} & \textbf{time (min.)}\\
\midrule
100 $\times$ \underline{10}	& \textcolor{orange} {Degenerate planar} & \textcolor{cyan} {14} & 15 \\
100 $\times$ \underline{25}	& \textcolor{orange} {Degenerate planar} & \textcolor{cyan} {14} & 45-60\\
100 $\times$ \underline{40}	& \textcolor{orange} {Degenerate planar} & \textcolor{cyan} {14} & 60-75\\
200 $\times$ \underline{10}	& \textcolor{orange} {Degenerate planar} & \textcolor{cyan} {14} & 30-45\\
200 $\times$ \underline{15}	& \textcolor{orange} {Degenerate planar} & \textcolor{cyan} {14} & 45\\
345 $\times$ \underline{10}	& \textcolor{orange} {Degenerate planar} & \textcolor{cyan} {14} & 45\\
100 $\times$ \underline{10}	& \textcolor{magenta} {Homeotropic} & \textcolor{cyan} {14} & 15\\
100 $\times$ \underline{25}	& \textcolor{magenta} {Homeotropic} & \textcolor{cyan} {14} & 60\\
100 $\times$ \underline{40}	& \textcolor{magenta} {Homeotropic} & \textcolor{cyan} {14} & 75\\
200 $\times$ \underline{10}	& \textcolor{magenta} {Homeotropic} & \textcolor{cyan} {14} & 30-45\\
200 $\times$ \underline{15}	& \textcolor{magenta} {Homeotropic} & \textcolor{cyan} {14} & 45-60\\
345 $\times$ \underline{10}	& \textcolor{magenta} {Homeotropic} & \textcolor{cyan} {14} & 30-45\\
200 $\times$ \underline{10}	& \textcolor{orange} {Degenerate planar} & 12 & 90\\
200 $\times$ \underline{10}	& \textcolor{magenta} {Homeotropic} & 12  & 90\\
200 $\times$ \underline{10}	& \textcolor{magenta} {Homeotropic} & \textcolor{green} {16} & 60\\
\bottomrule
\end{tabular}
\end{table}
We conclude that silane derivatives~($e.g.$, DMOAP), usually employed for generating homeotropic alignment for thermotropic LCs, may not be effective for inducing normal surface anchoring of the LCLCs. Based on previous report by Tone \textit{et al.}, both hydrophobicity and chemical affinity could play a role in inducing homeotropic alignment. Here, materials with low surface energy--without alkyl chains--could be considered as suitable alternatives to generate homeotropic alignment, for instance polybutadiene and styrene-butadiene-styrene \cite{Tone2013}. These materials can work for glass devices, however there are a few technical hurdles in coating PDMS-glass microchannels with these materials, primarily due to the dissolution step (with organic solvents) which can cause swelling of the PDMS, thereby modifying the microfluidic geometry.

\begin{figure}
\centering
\includegraphics[width=12cm]{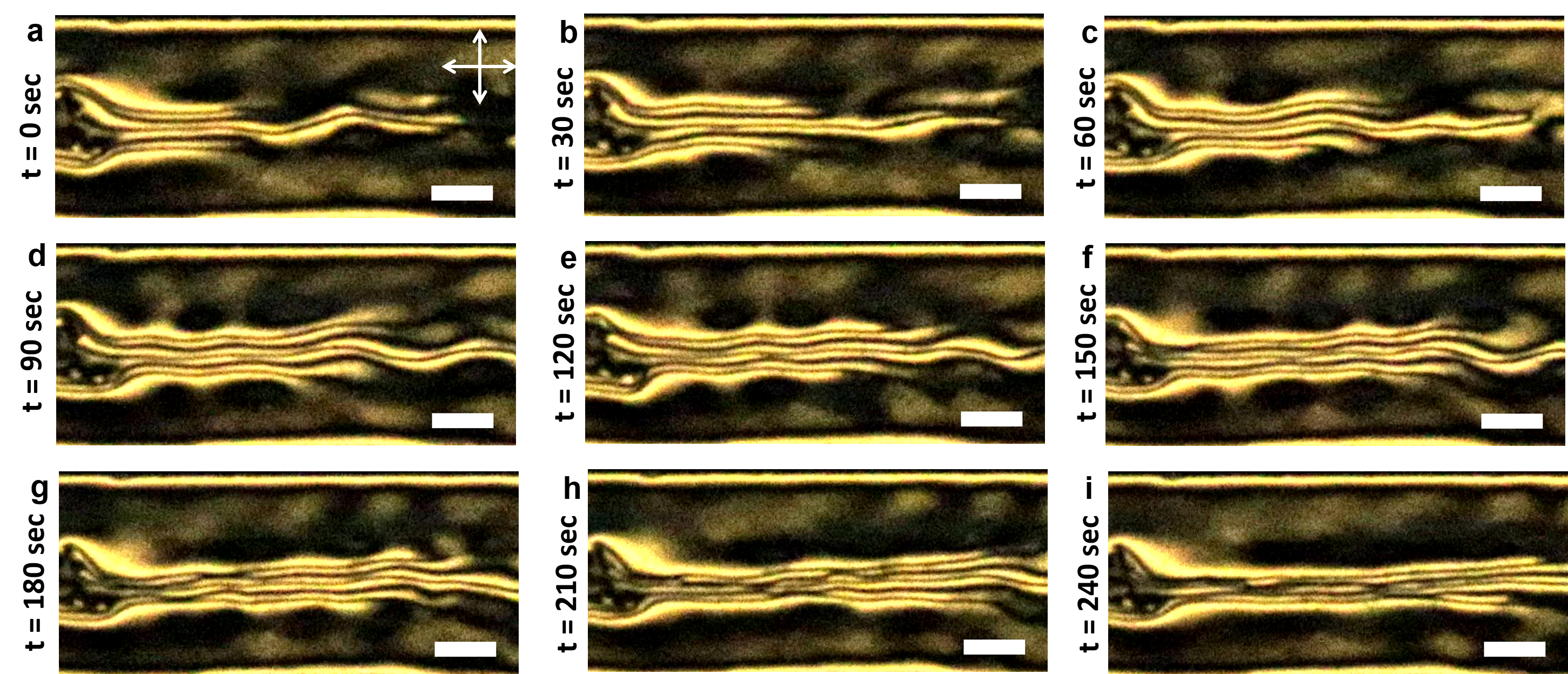}
\caption{\label{fig:Figure7-flowDMOAP}\textbf{Psuedo flow textures exhibited by the N phase of 14~wt.~\% DSCG between crossed polarizers}, imaged within a 345 $\mu$m wide, 10 $\mu$m deep microchannel with homeotropic anchoring. (\textbf{a})-(\textbf{1}) time evolution of disclination lines. Scale bar: 100 $\mu$m.}
\end{figure}

The key and the most striking result of this work is the temporal variation of lyotropic textures induced by a spontaneous phase transition from nematic to the M-phase. By changing the channel dimensions, surface anchoring, and the concentration of the LCLC material, we could systematically tune both the nature of the characteristic texture and the temporal dynamics of the nematic-to M-phase transition. Additionally, the texture of the M-phase depended on the surface anchoring conditions: herringbone texture under degenerate planar conditions; and spherulite texture under homeotropic conditions. Furthermore, both herringbone and spherulite textures manifested global orientations: while the herringbone texture was oriented along the channel length, the spherulite texture was oriented perpendicular to the channel length (in the transverse direction). The temporal dynamics of the nematic-M-phase transition could be controlled by varying the channel width (larger width lead to longer transition time), whereas changing the depth primarily impacted the textural uniformity. Our straightforward strategy to employ LCLC self-organization within soft-lithography based microfluidic platform considerably widens the scope of LCLCs, for both applied and fundamental research, and sets the stage for future studies on the flow behaviour LCLCs in $strict$ microfluidic environments. Interestingly, even the static $no$-$flow$ textures hold promising prospects for potential applications in LCLC-mediated transport processes. As shown through time lapse image sequence in Figure\ref{fig:Figure7-flowDMOAP}, defect lines near a microscale inclusion (channel and inclusion were DMOAP-treated) can transition over time--purely due to surface-mediated effects--thereby triggering microscale transport, even in absence of external pressure gradients~\cite{Sharma2020Surface-mediatedPrepration}. Investigations on the dynamics of LCLCs under flow within PDMS-glass microfluidic devices is currently underway, and can be extended to other systems to study surface-mediated phase transitions, for instance in water based lyotropic liquid crystals systems such as peptides, liquid crystal polymers, and viruses. The temporal stability of textures can be of significant implication on real-world applications, particularly to capture the dynamics of in biological systems, where transport processes can respond to (or be modulated by) the compliance, surface and morphological attributes of the system.

\section{Author contributions}
A. Sharma and A. Sengupta designed research, performed experiments and analyzed data. All authors interpreted the data and wrote the paper.
\section{Acknowledgments}
This work was supported by the ATTRACT Investigator Grant (A17/MS/11572821/MBRACE) of the Luxembourg National Research Fund. The authors are grateful to M. G. Mazza, A. Ghoshal and J. Dhar for valuable discussions during the course of this work.

\bibliographystyle{unsrt}  
\bibliography{references1}  
\section{Supplementary figures}
\begin{figure}
\centering
\includegraphics[width=6cm]{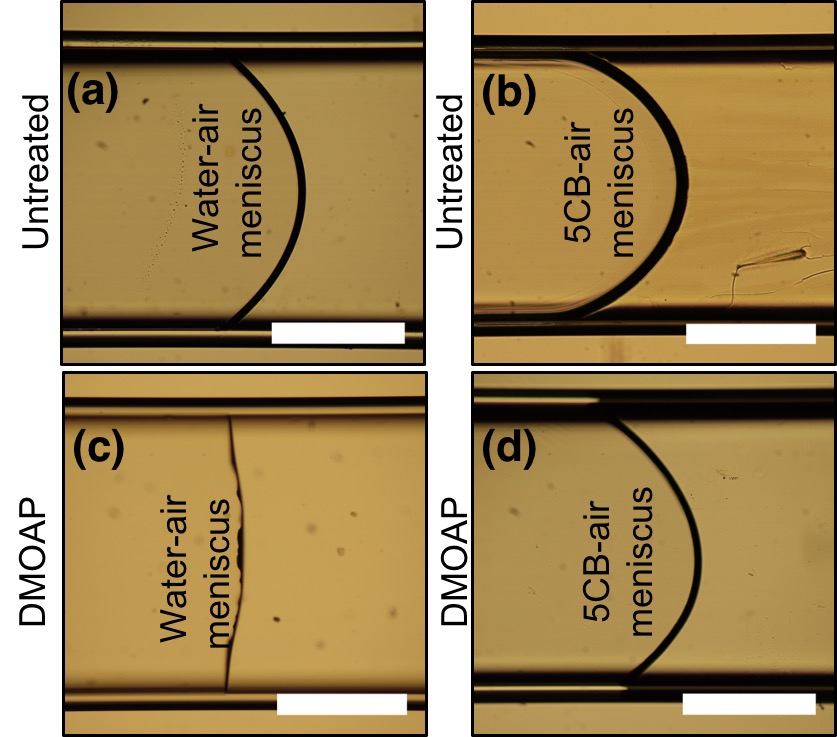}
\caption{\label{fig:Meniscus}\textbf{Optical micrographs of menisci within glass capillaries with dimensions 100 $\times$ 1000~$\mu$m (cross-section dimensions)}, without any treatment: (\textbf{a}) Water; (\textbf{b}) 5CB (filled in nematic phase). Optical micrograph of glass capillary after DMOAP treatment: (\textbf{c}) Water and (\textbf{d}) 5CB (filled in isotropic  phase). Scale bar: 500 $\mu$m.}
\end{figure}

\begin{figure}
\centering
\includegraphics[width=10cm]{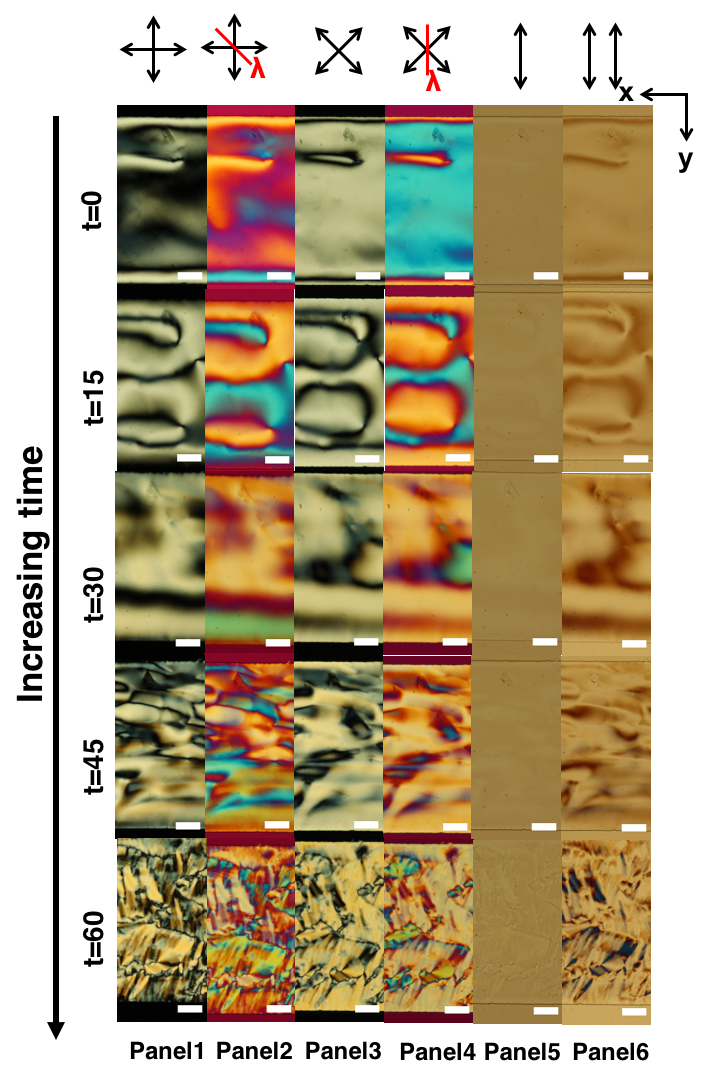}
\caption{\label{fig:plasma345-10-1}\textbf{Degenerate planar anchoring within wider channels}, here using 345$\mu$m wide, 10 $\mu$m deep microfluidic devic. The textural transition to herringbone texture is accelerated in wider channels, as captured by the time lapse imaging with polaizers at various configureations:(\textbf{Panel 1}) between crossed polarizers; (\textbf{Panel 2}) crossed polarizers with the $\lambda$-plate; (\textbf{Panel 3}) crossed polarizers with channel at $45^\circ$ relative to the polarizer; and with $\lambda$-plate (\textbf{Panel 4}); (\textbf{Panel 5}) with analyzer only and (\textbf{Panel 6}) analyzer parallel to the polarizer, both perpendicular to the channel length. Here, time (t) is in minutes after the cooling ramp is completed). Scale bar: 50 $\mu$m.}
\end{figure}

\begin{figure}
\centering
\includegraphics[width=12cm]{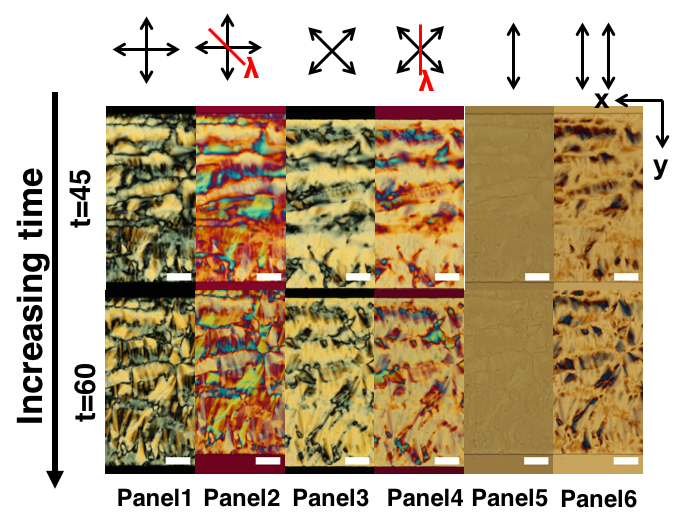}
\caption{\label{fig:plasma345-10-2}\textbf{Herringbone texture in due to degenerate planar anchoring within 345$\mu$m wide, 10$\mu$m deep microchannel.} The POM images were taken using the same protocol as described above. Scale bar: 50 $\mu$m}
\end{figure}

\begin{figure}
\centering
\includegraphics[width=15cm]{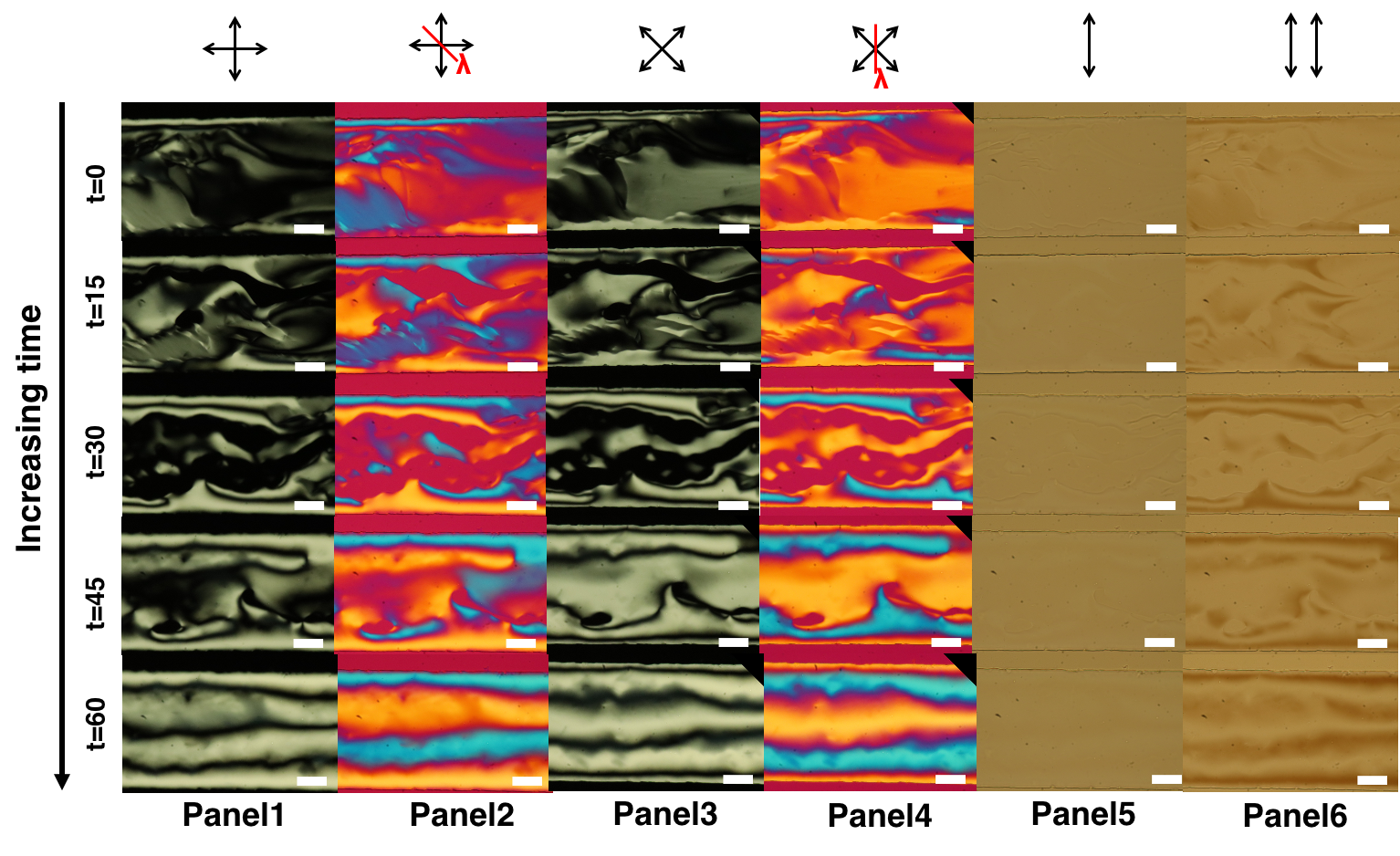}
\caption{\label{fig:DMOAP200-10-1} \textbf{Development of textures in 200$\mu$m wide, 10$\mu$m deep microchannel possessing degenerate planar anchoring} using 14~wt.~\% DSCG. The textural evolution was observed as a function of time, using polarizing optical microscopy oover a time span of 60 minutes. The configuration of the polarizer and $\lambda$-retardation plate are presented at the top of each panel. Here, time (t) is in minutes after cooling ramp is concluded). Scale bar: 50 $\mu$m.}
\end{figure}

\begin{figure}
\centering
\includegraphics[width=15cm]{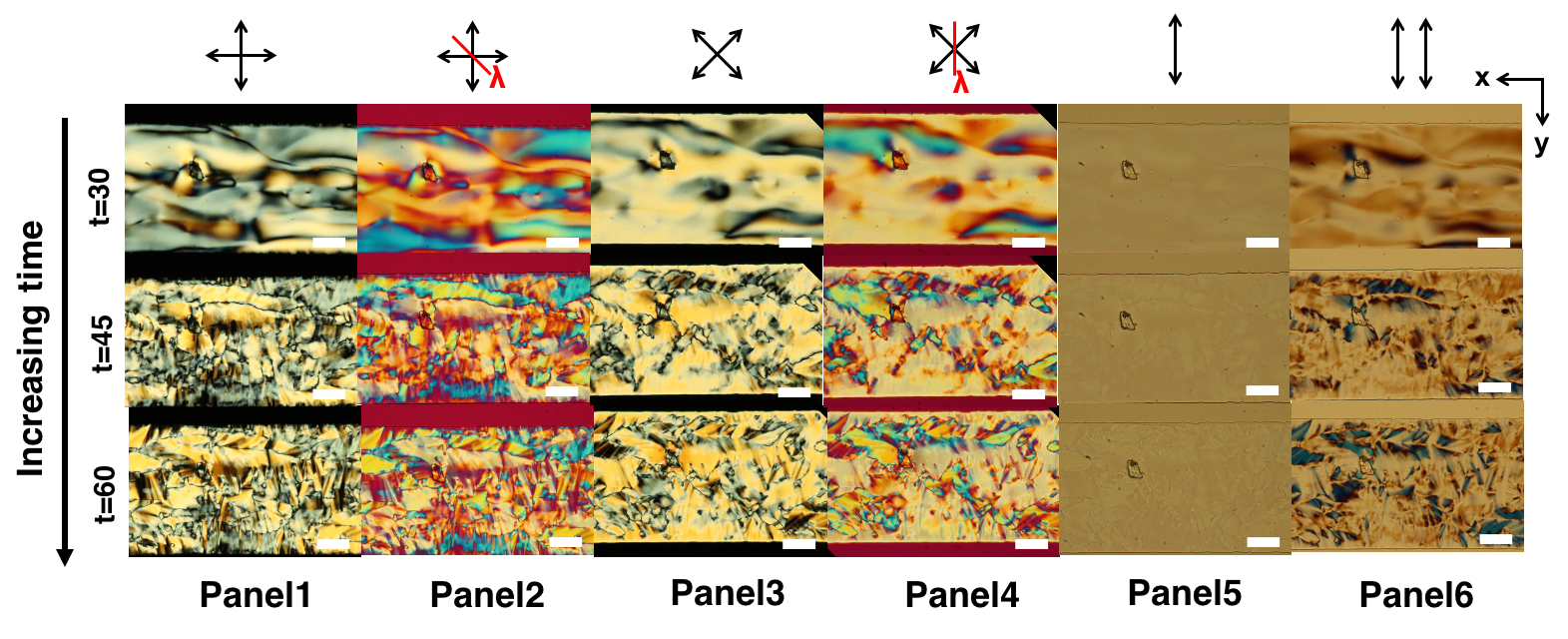}
\caption{\label{fig:DMOAP200-10-2}\textbf{Long term textures in 200$\mu$m wide, 10$\mu$m deep microchannels treated for hometropic anchoring.}. The nucleation and propagation of M-phase can be observed here, imaged between polarizers under different configurations as described above. Scale bar: 50 $\mu$m.}
\end{figure}

\begin{figure}
\centering
\includegraphics[width=15cm]{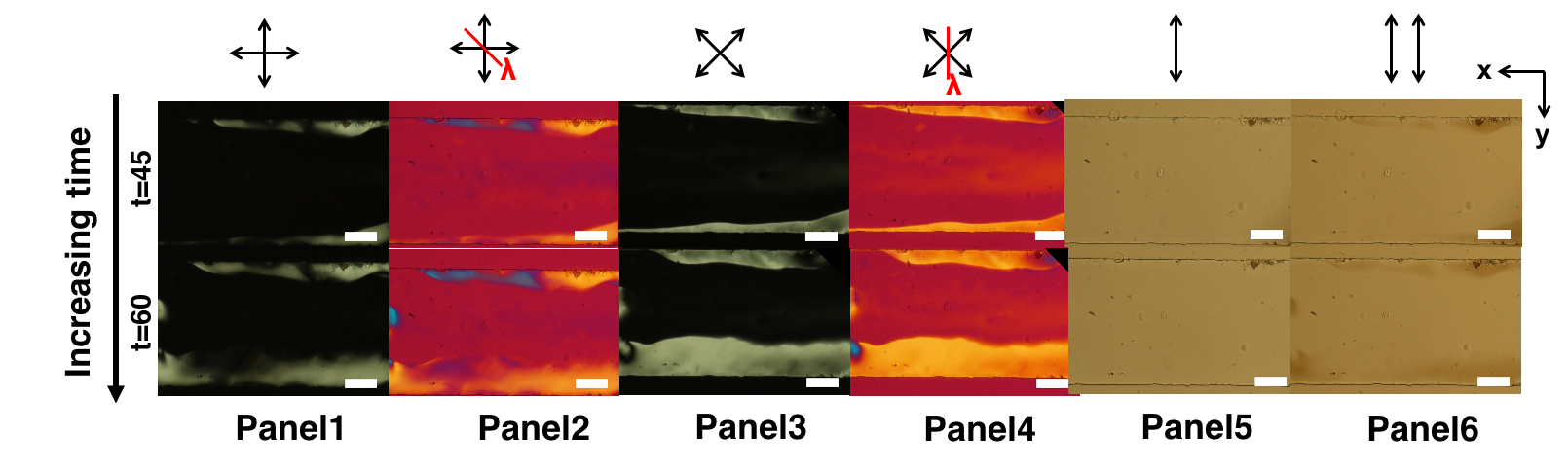}
\caption{\label{fig:DMOAP200-10-3}\textbf{Stable homeotropic anchoring in certain regions within the DMOAP-treated 200$\mu$m wide, 10$\mu$m deep microchannel}, with 14~wt.~\% DSCG solution. The textures were captured between polarizers under different configurations, as described above. Scale bar: 50 $\mu$m.}
\end{figure}

\begin{figure}
\centering
\includegraphics[width=10cm]{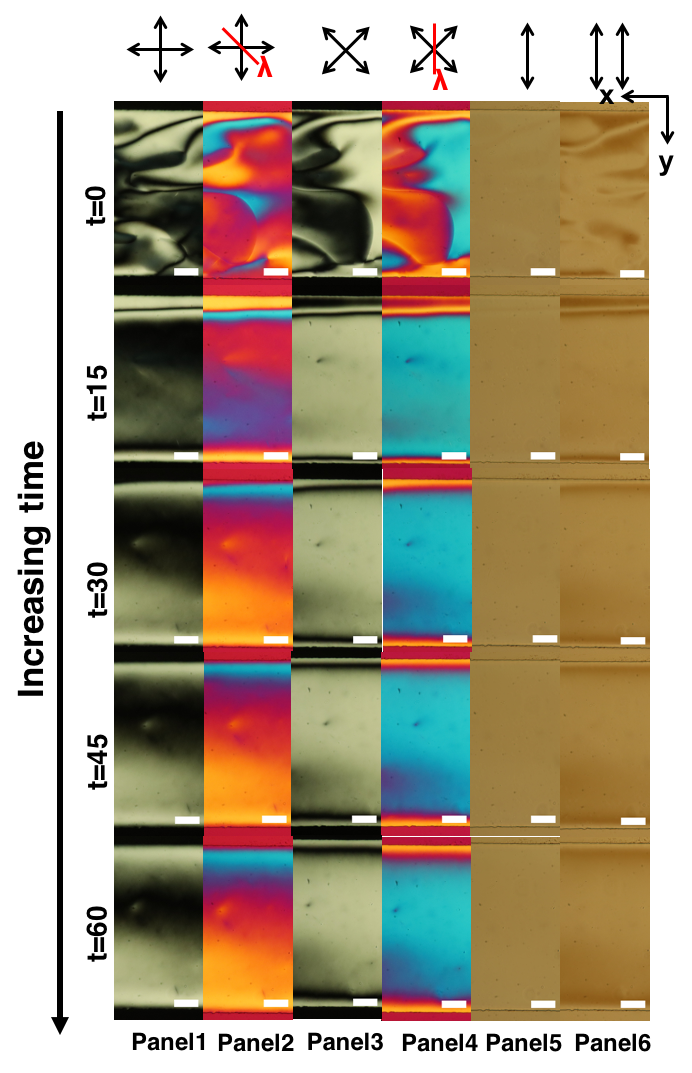}
\caption{\label{fig:DMOAP345-10-1}\textbf{Homeotropic anchoring in wider channels, here within 345$\mu$m wide, 10$\mu$m deep microchannel} lead to the $pseudo$ $planar$ texture, with 14~wt.~\% DSCG. Overall, the texture resembled that of planar or tilted orientation. Imaging details are noted at the top of each panel. Scale bar: 50 $\mu$m.}
\end{figure}

\begin{figure}
\centering
\includegraphics[width=12cm]{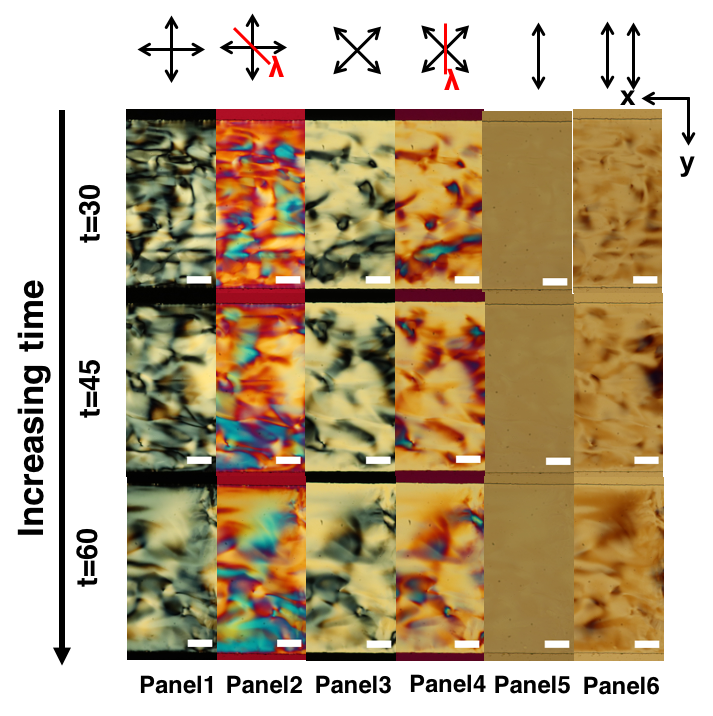}
\caption{\label{fig:DMOAP345-10-2}\textbf{Gradual emergence of the spherulite texture} within DMOAP-treated microschannel (345$\mu$m wide, 10$\mu$m deep), with 14~wt.~\% DSCG. Imaging started at 30 minutes after the end of the cooling ramp, and lasted till 60 minutes: (\textbf{Panel 1}) between crossed polarizers; (\textbf{Panel 2}) crossed polarizers and $\lambda$-plate; (\textbf{Panel 3}) crossed polarizers at $45^\circ$ relative to channel length; (\textbf{Panel 4}) under crossed polarizers with $45^\circ$ rotation and $\lambda$-plate; (\textbf{Panel 5}) with only analyzer and (\textbf{Panel 6}) analyzer parallel to the polarizer, with both perpendicular to the channel length). Scale bar: 50 $\mu$m.}
\end{figure}

\begin{figure}
\centering
\includegraphics[width=14cm]{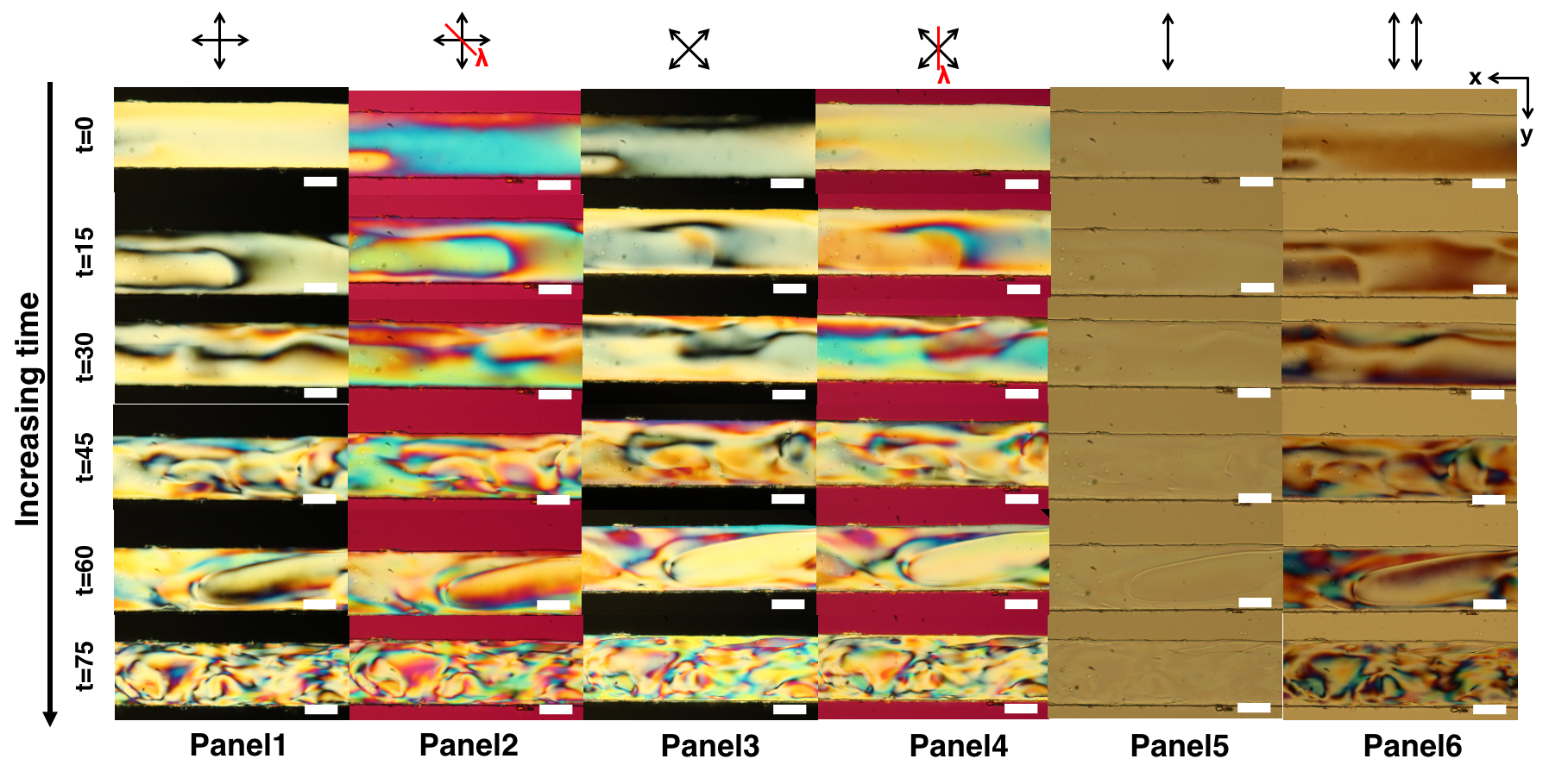}
\caption{\label{fig:plasma-100-25-1}\textbf{Degenerate planar anchoring-(100$\mu$m wide, 25$\mu$m deep)-14~wt.~\% DSCG.} Evolution of textures exhibited as a function of time under various configurations of the polarizers. Polarizing optical micrographs show the top view of the channel from t=0 to 75 minutes, images were taken as per the scheme discussed in the previous images. Here, time (t) is in minutes after ramp is completed. Scale bar: 50 $\mu$m.}
\end{figure}

\begin{figure}
\centering
\includegraphics[width=14cm]{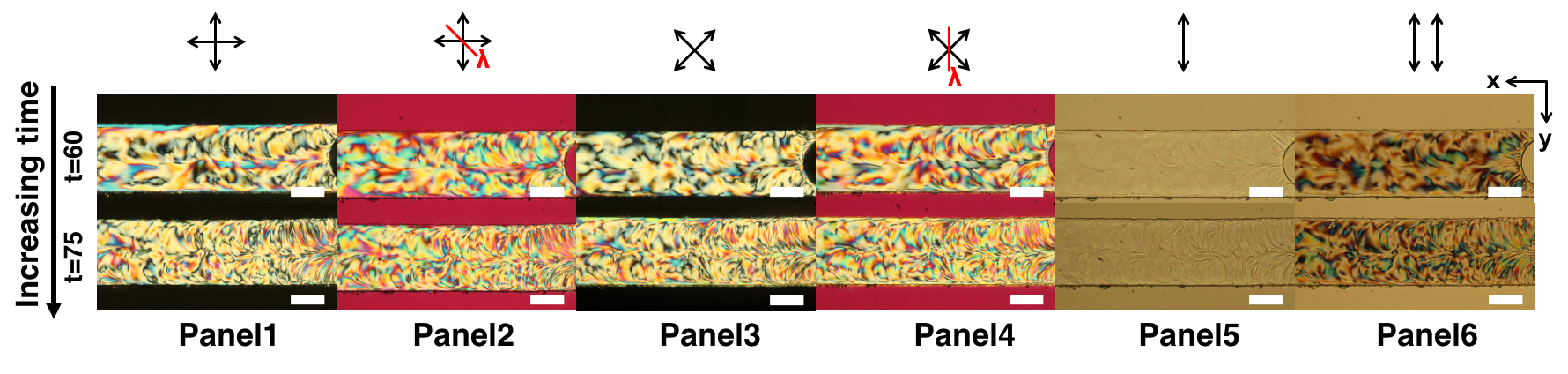}
\caption{\label{fig:plasma-100-25-2} \textbf{M-phase texture in 100$\mu$m wide, 25$\mu$m deep microchannel.} Evolution of textures exhibited as a function of time under various conditions. Polarizing optical micrographs showing the top view of the channel from t=60 to 75 minutes. Evolution of textures under crossed polarizers; images were taken as per the scheme discussed in the previous images. Here, time (t) is in minutes after ramp is completed. Scale bar: 50 $\mu$m.}
\end{figure}

\begin{figure}
\centering
\includegraphics[width=14cm]{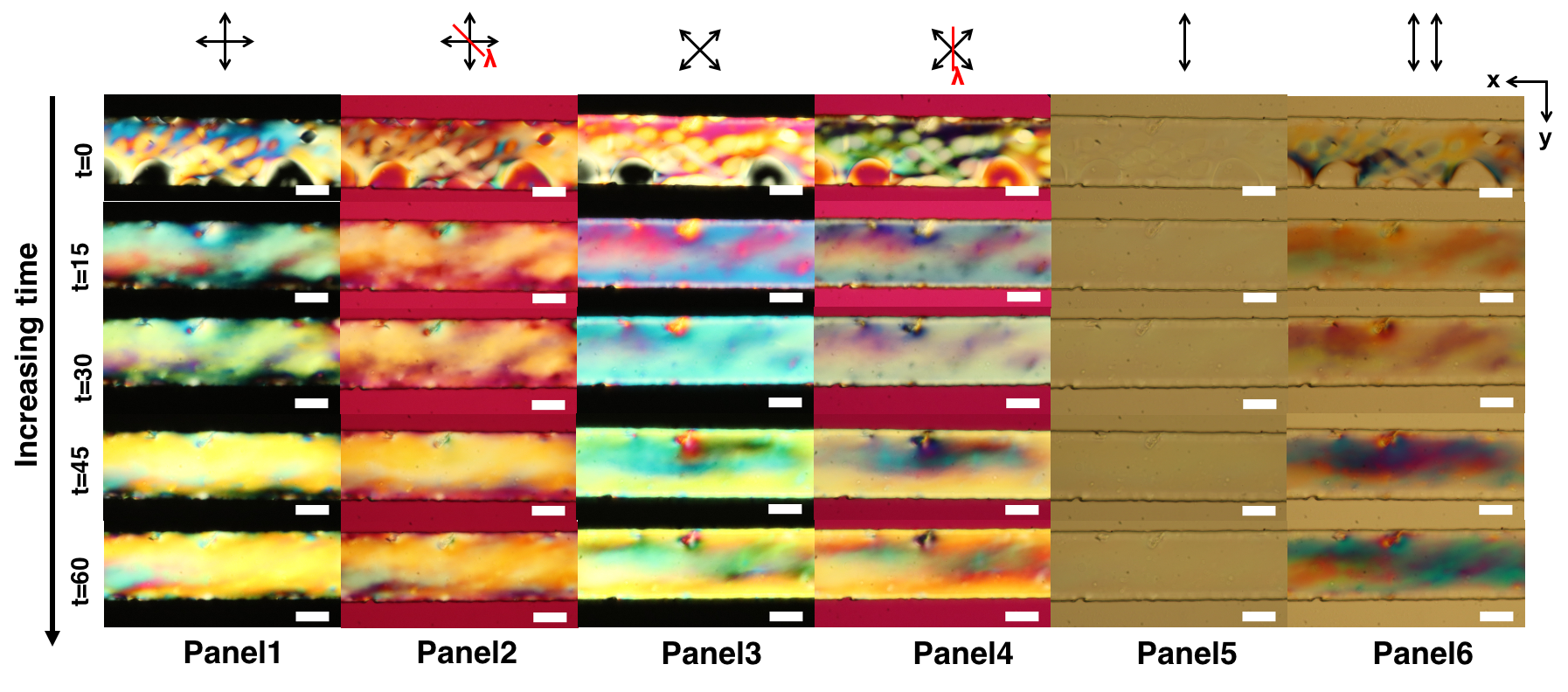}
\caption{\label{fig:plasma-100-40-1} \textbf{Degnerate planar anchoring textures in 100$\mu$m wide, 40$\mu$m deep channel}. Evolution of textures as a function of time under various polarizer configurations, as described in previous images. Scale bar: 50 $\mu$m.}
\end{figure}

\begin{figure}
\centering
\includegraphics[width=14cm]{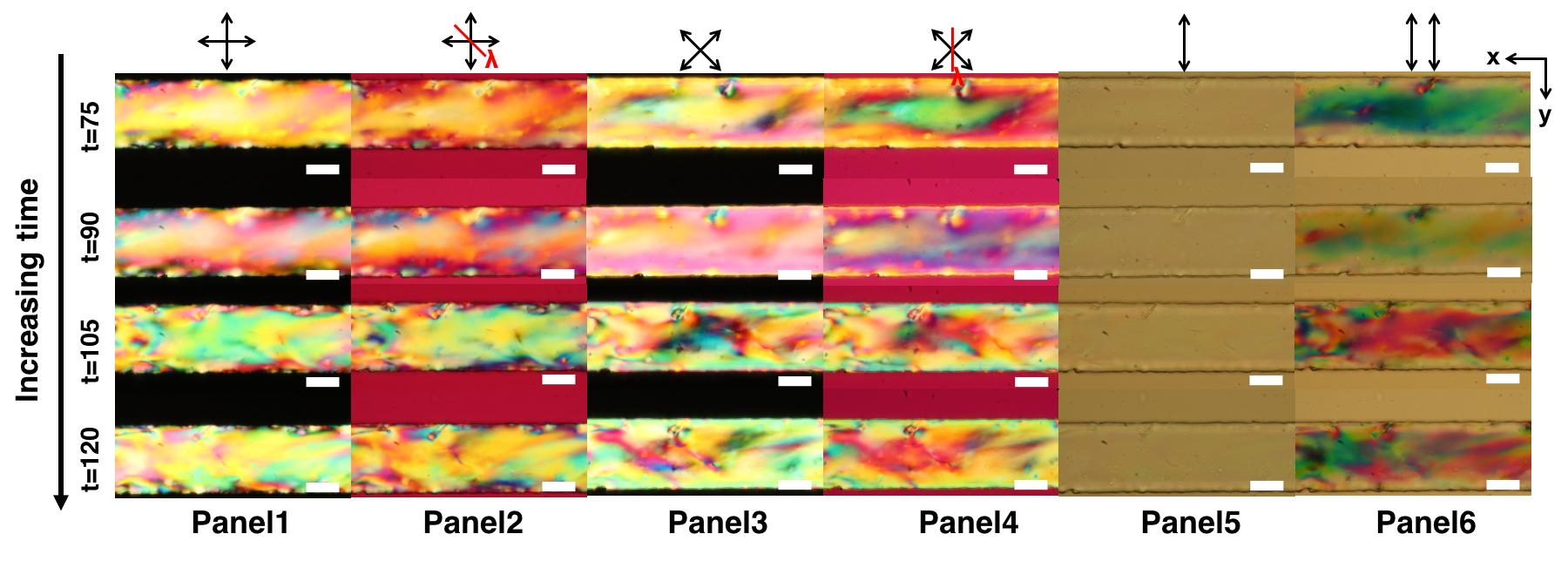}
\caption{\label{fig:plasma-100-40-2} \textbf{Long term textures in 100$\mu$m wide, 40$\mu$m deep channels.} The absence of the well-aligned herringbone texture, under degenerate planar boundary conditions, suggests that the textural transitions are hindered within $deeper$ channels. Scale bar: 50 $\mu$m.}
\end{figure}

\begin{figure}
\centering
\includegraphics[width=15cm]{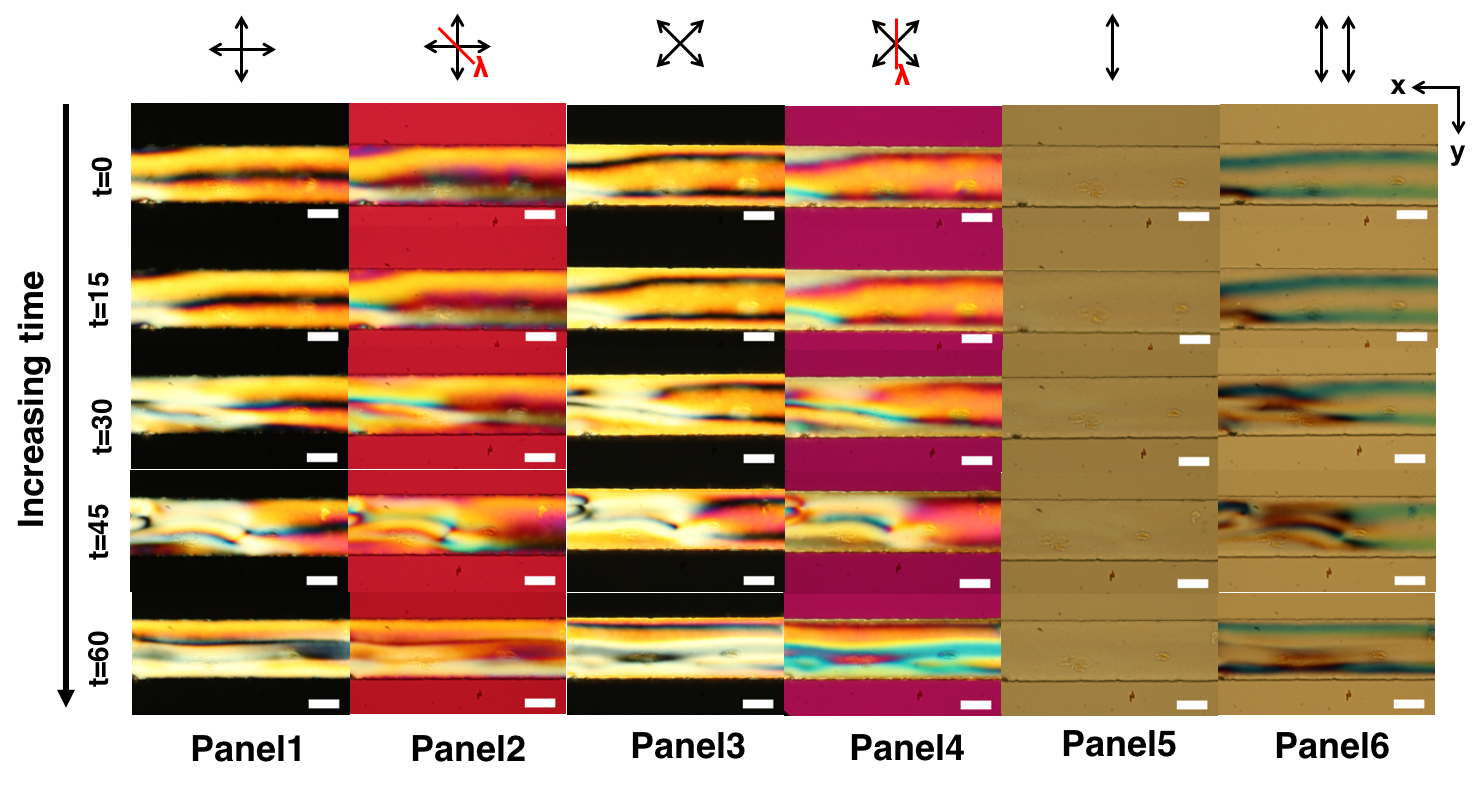}
\caption{\label{fig:DMOAP-100-25-1}\textbf{Effect of the channel depth on the DSCG texture}, here observed within 100$\mu$m wide, 15$\mu$m deep microchannel possessing homeotropic anchoring conditions. The concentration of DSCG was 14~wt.~\%. Polarizing optical micrographs were taken using same configuration of polarizers and $\lambda$-plate, as described previously. Scale bar: 50 $\mu$m.}
\end{figure}

\begin{figure}
\centering
\includegraphics[width=15cm]{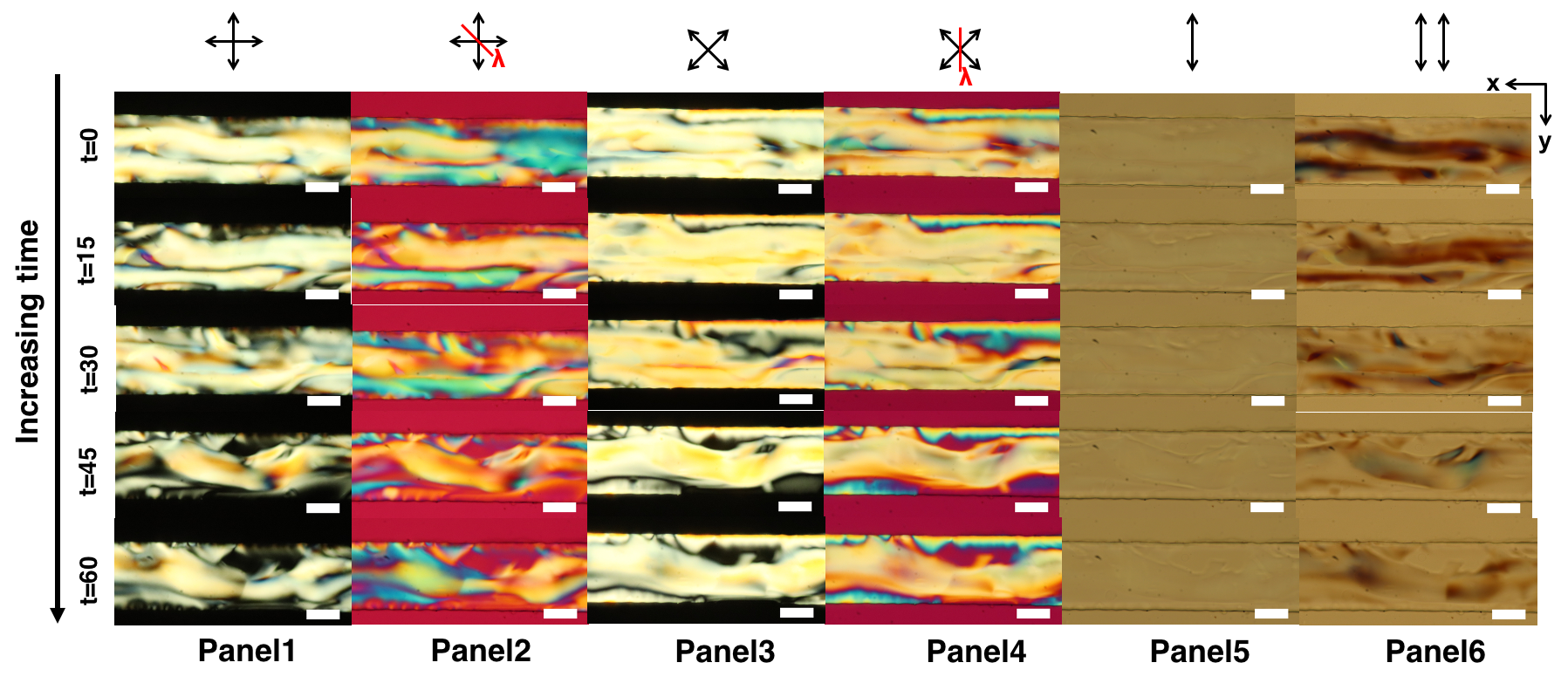}
\caption{\label{fig:DMOAP-100-40-1}\textbf{Effect of the channel depth on the DSCG texture}, here observed within 100$\mu$m wide, 40$\mu$m deep microchannel, using the 14~wt.~\% DSCG. Imaging was done following details described previously. Scale bar: 50 $\mu$m.}
\end{figure}

\begin{figure}
\centering
\includegraphics[width=14cm]{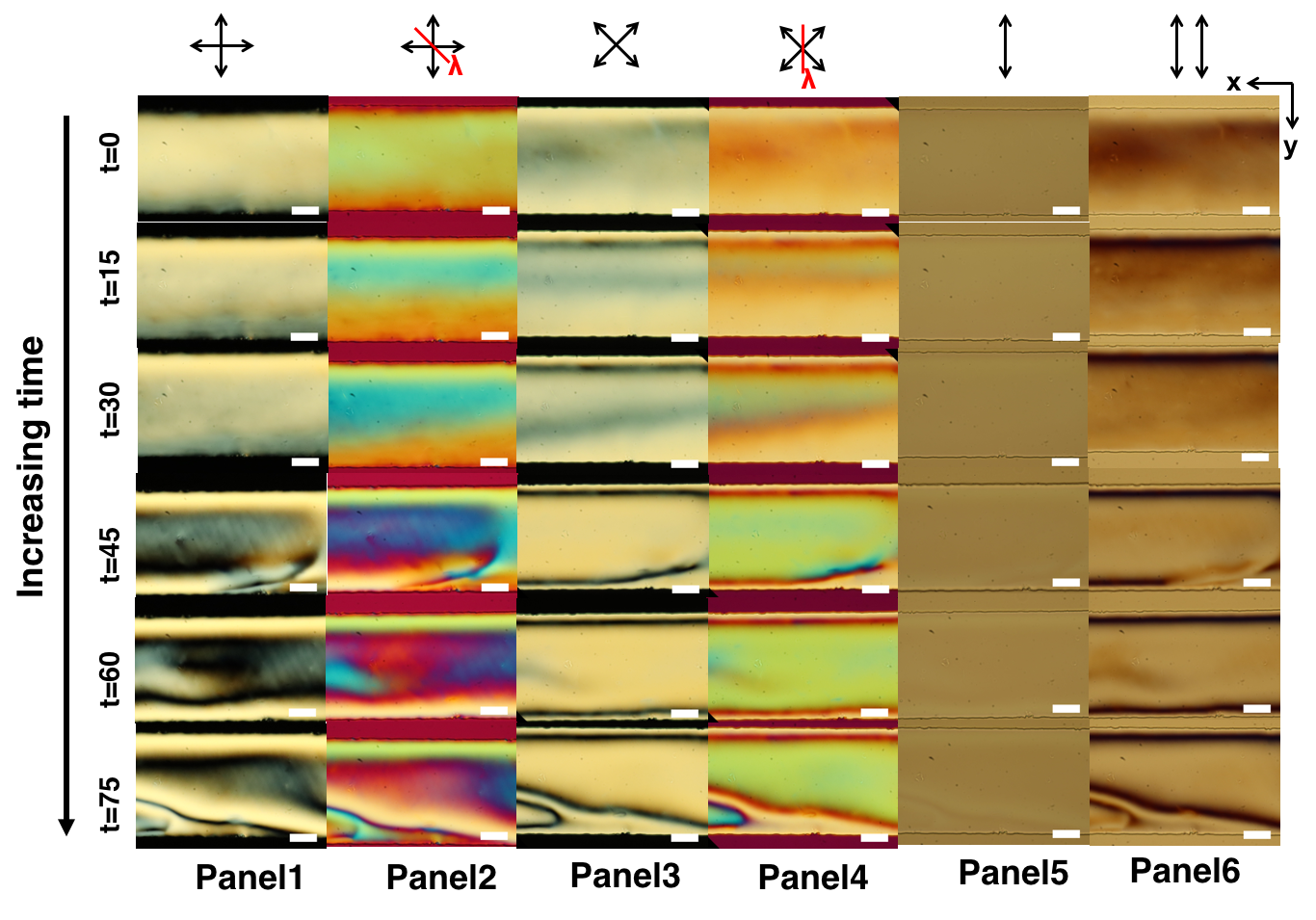}
\caption{\label{fig:plasma-200-15-1}\textbf{DSCG textures within 200$\mu$m wide, 15$\mu$m deep microchannel possessing degnerate planar anchoring}, here observed with 14~wt.~\% DSCG. Scale bar: 50 $\mu$m.}
\end{figure}

\begin{figure}
\centering
\includegraphics[width=14cm]{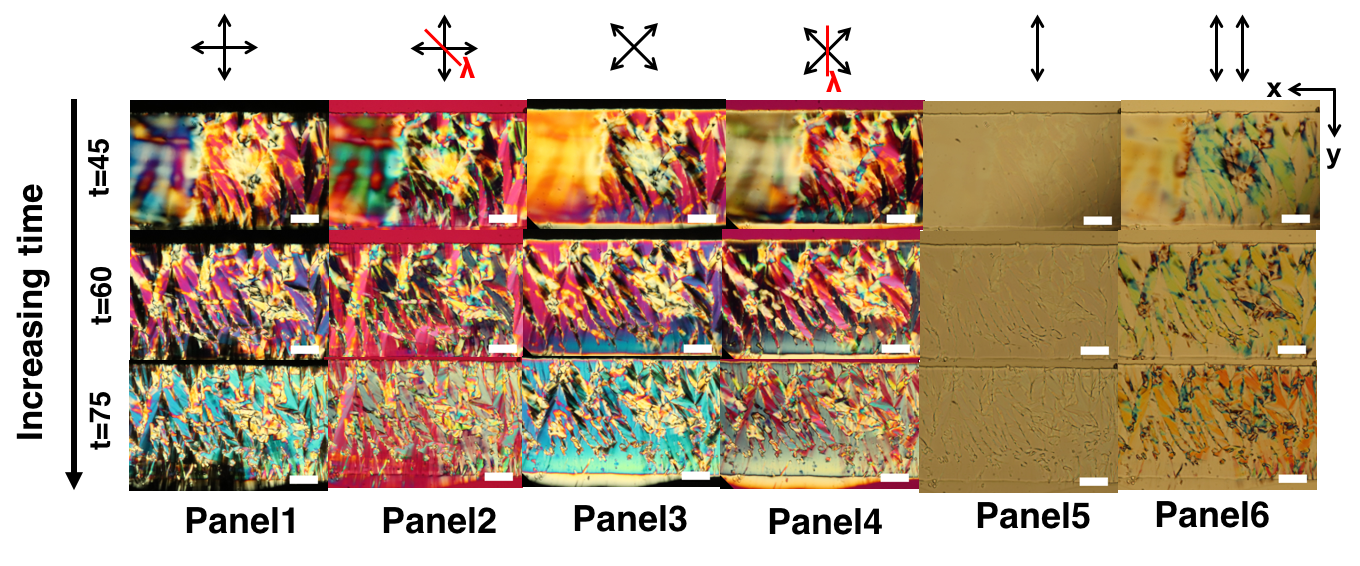}
\caption{\label{fig:plasma-200-15-2}\textbf{Degnerate planar anchoring within 200$\mu$m wide, 15$\mu$m deep microchannel} yields M-phase textures at 14~wt.~\% DSCG. Imaging details are described previously. Scale bar: 50 $\mu$m.}
\end{figure}

\begin{figure}
\centering
\includegraphics[width=15cm]{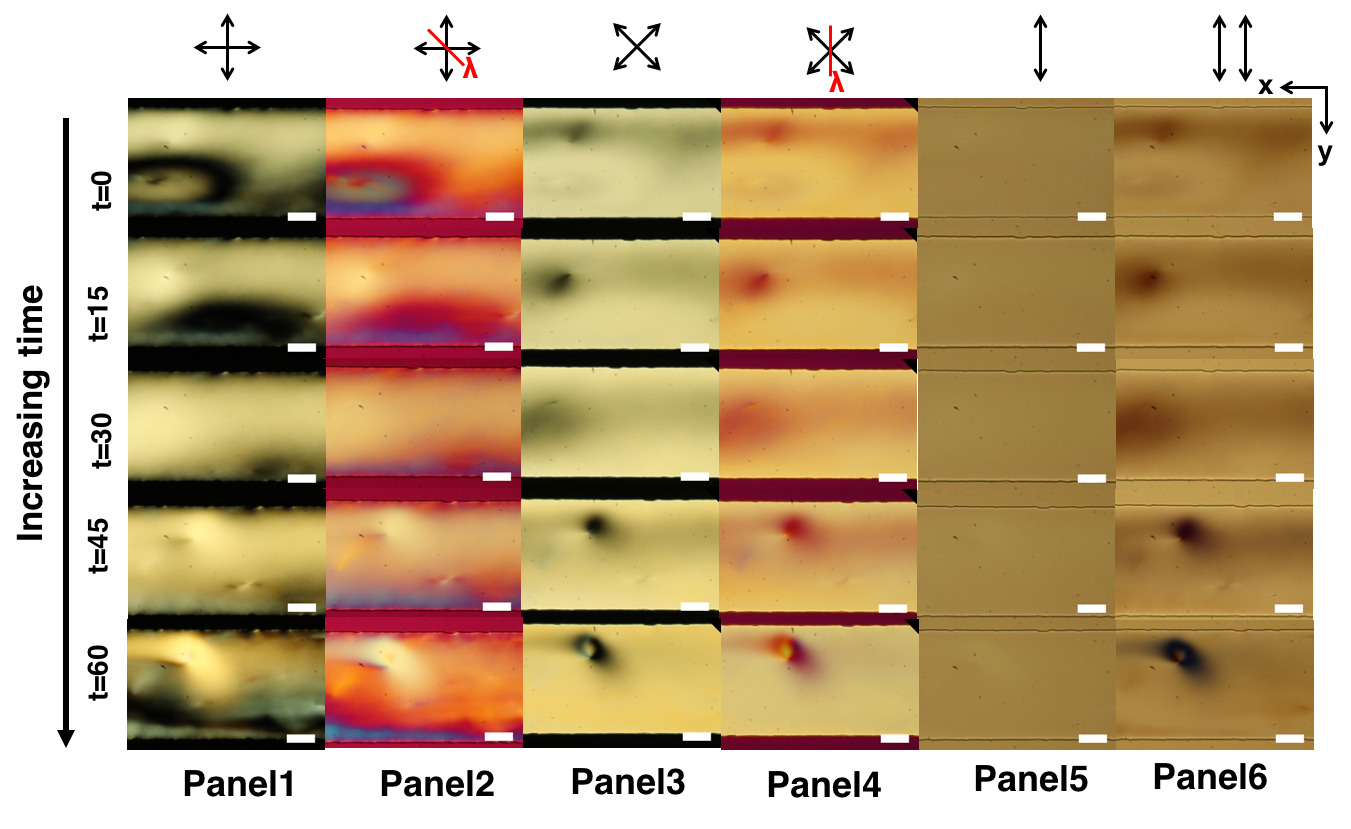}
\caption{\label{fig:DMOAP-200-15-1}\textbf{DSCG textures within 200$\mu$m wide, 15$\mu$m deep microchannel possessing homeotropic surface anchoring}, here shown for 14~wt.~\% DSCG. Scale bar: 50 $\mu$m.}
\end{figure}

\begin{figure}
\centering
\includegraphics[width=15cm]{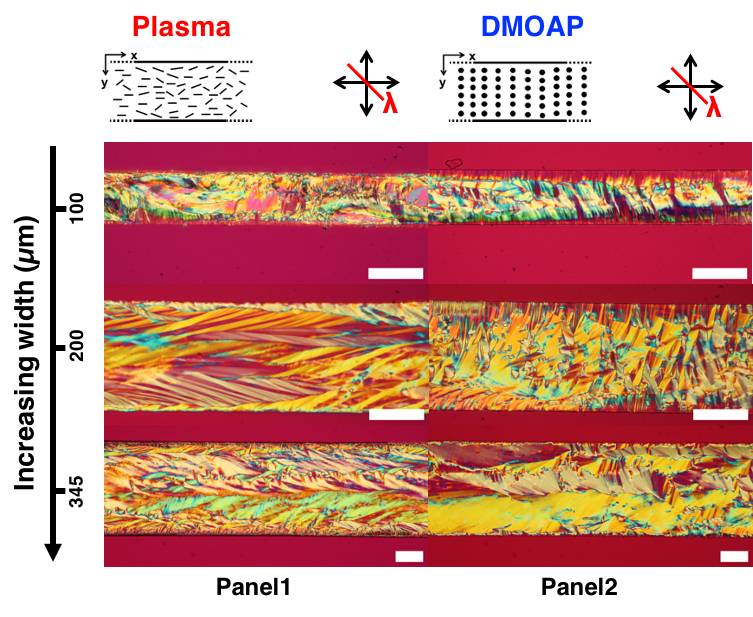}
\caption{\label{fig:M-phase-all}\textbf{Herringbone and spherulite microfluidic textures formed by DSCG under degenerate planar and homeotropic anchoring conditions respectively.} The textures corresponding to DSCG M-phase were observed between crossed polarizers with $\lambda$-retardation plate. The textures depend on the channel aspect ratio and imposed surface anchoring. (\textbf{Panel 1}) Microchannel with degnerate planar anchoring with dimensions: (100$\mu$m wide, 10$\mu$m deep), (200$\mu$m wide, 10$\mu$m deep) and (345$\mu$m wide, 10$\mu$m deep), respectively. Note the structural orientation of the herringbone texture along the channel length. (\textbf{Panel 2}) Observation of the spherulite texture in microchannels possessing homeotropic anchoring. The spherulite texture exhibits an overall orientation along the channel transverse direction (wall-to-wall). Scale bar: 100 $\mu$m.}
\end{figure}

\begin{figure}
\centering
\includegraphics[width=12cm]{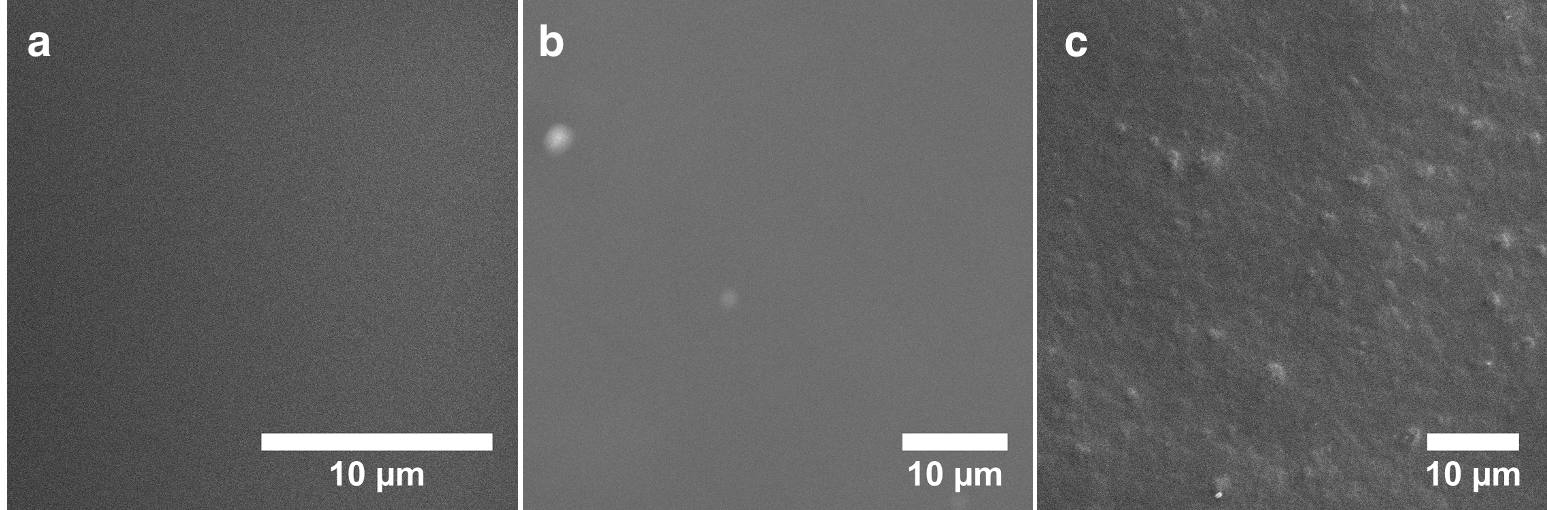}
\caption{\label{fig:SEM}\textbf{SEM of PDMS under the two different anchoring conditions}: (\textbf{a}) Pristine, untreated PDMS surface; and (\textbf{b}) Plasma-treated PDMS; and (\textbf{c}) DMOAP-treated PDMS surface.}
\end{figure}

\begin{figure}
\centering
\includegraphics[width=10cm]{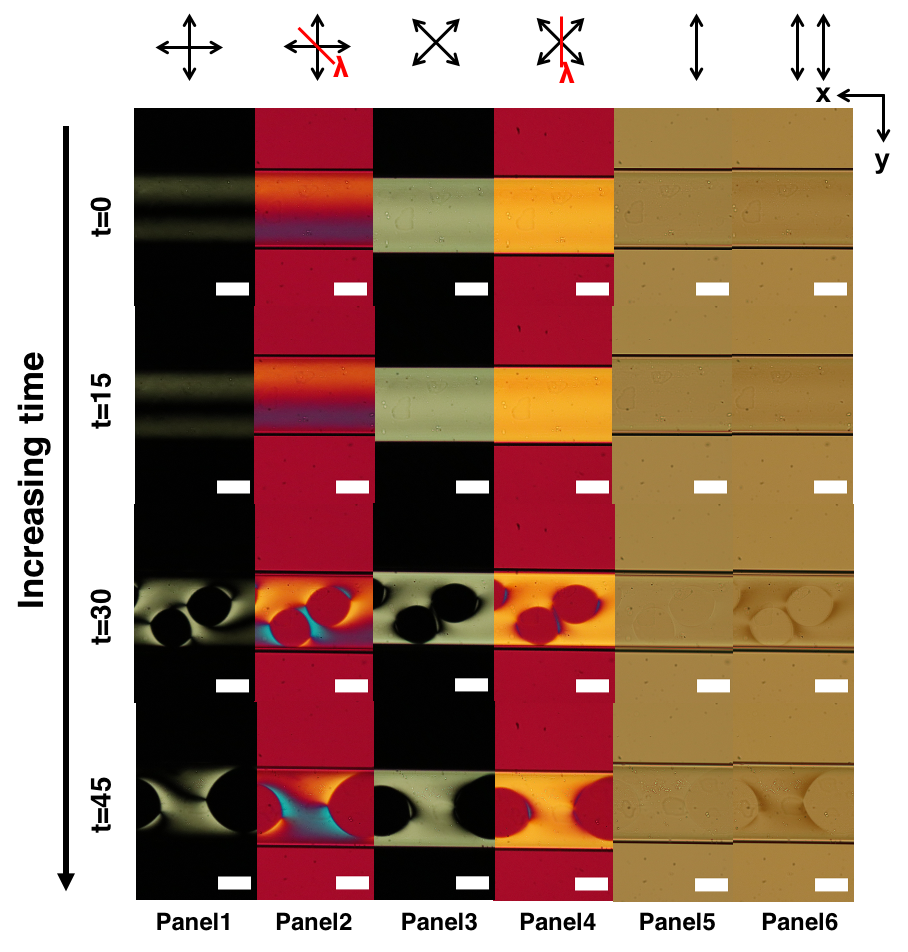}
\caption{\label{fig:POM14DSCGcapillarynotreatment}\textbf{DSCG textures within untreated glass capillaries.} Evolution of textures exhibited by 14~wt.~\% DSCG in 10 $\times$ 100~$\mu$m (height $\times$ width) glass capillary as a function of time. Polarizing optical micrographs show the top view of the channel from t=0 to 45 minutes: (\textbf{Panel 1}) between crossed polarizers; (\textbf{Panel 2}) crossed polarizers with $\lambda$-plate; (\textbf{Panel 3}) crossed polarizers at $45^\circ$ relative to channel length; and with $\lambda$-plate (\textbf{Panel 4}); (\textbf{Panel 5}) analyzer only and (\textbf{Panel 6}) parallel analyser and polarizer. Scale bar: 50 $\mu$m.}
\end{figure}

\begin{figure}
\centering
\includegraphics[width=10cm]{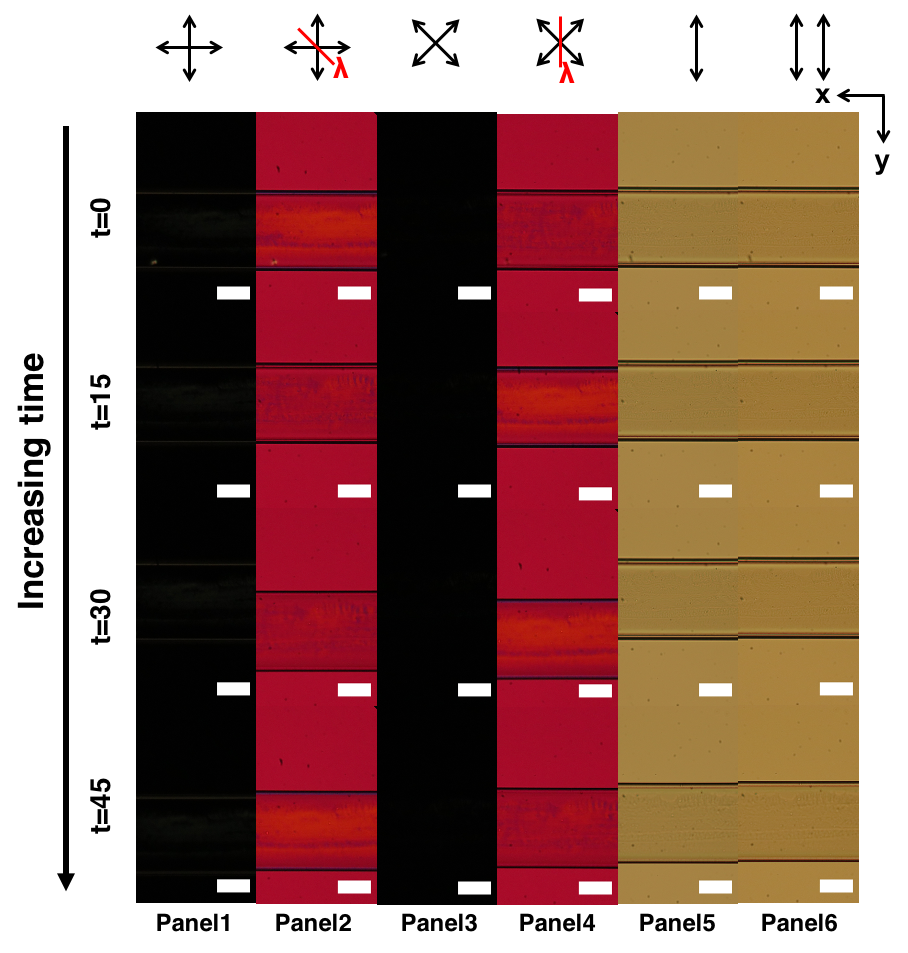}
\caption{\label{fig:POM14DSCGcapillaryDMOAP}\textbf{DSCG textures within DMOAP-treated glass capillaries.} Evolution of textures exhibited by 14~wt.~\% DSCG in 10 $\times$ 100~$\mu$m (height $\times$ width) DMOAP-treated glass capillary as a function of time. POM shows the top view of the channel from t=0 to 45 minutes, imaging was carried out as per the protocol discussed above. Scale bar: 50 $\mu$m.}
\end{figure}

\begin{figure}
\centering
\includegraphics[width=14cm]{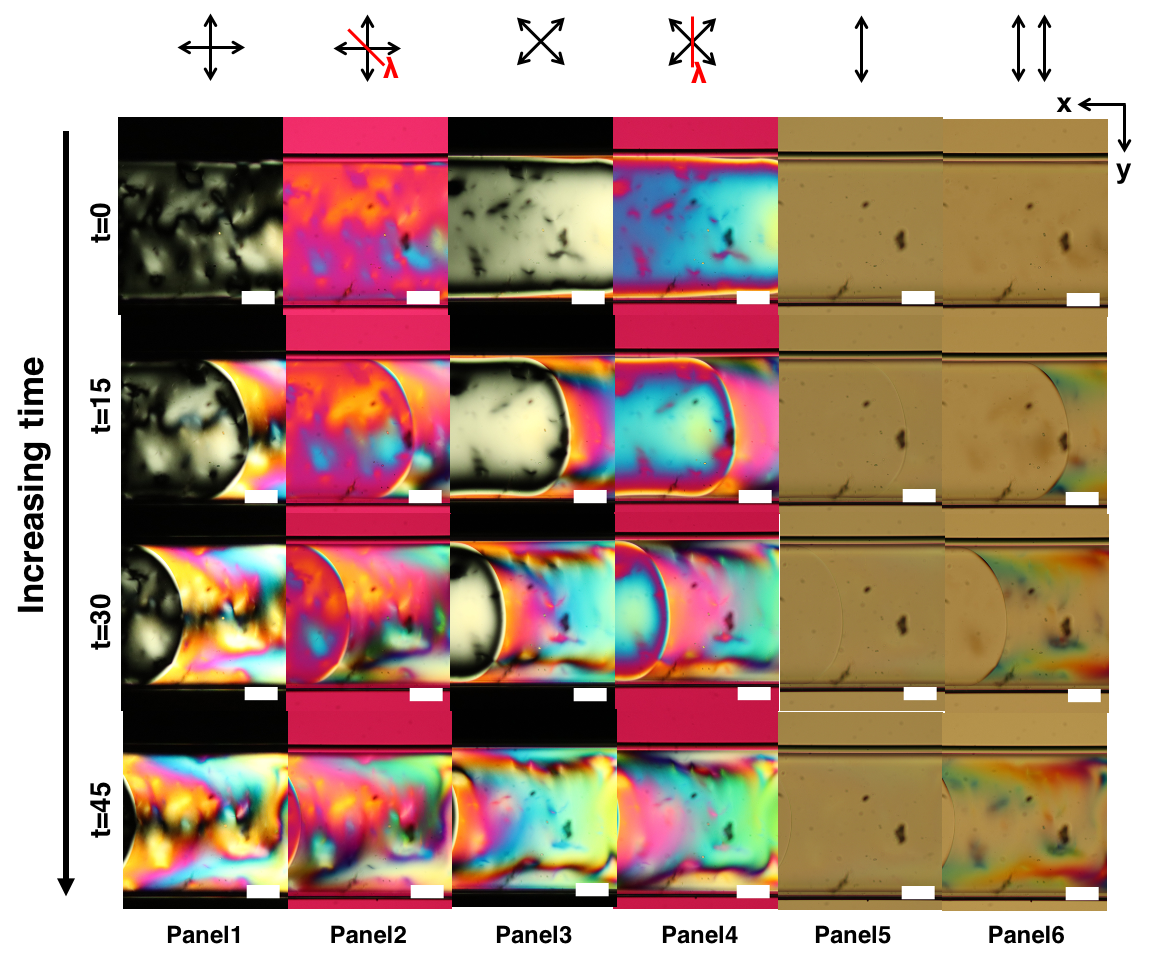}
\caption{\label{fig:Openglasscapillary}\textbf{Evolution of DSCG textures (14~wt.~\% DSCG) in 40 $\times$ 400~$\mu$m untreated glass capillary, open from both sides}. The transition from nematic to M-phase could be reproduced by allowing evaporation of water from the original DSCG solution (14~wt.~\%). The sequence of images follow similar polarization configuration as described throughout the text. Scale bar: 100 $\mu$m.}
\end{figure}

\begin{figure}
\centering
\includegraphics[width=14cm]{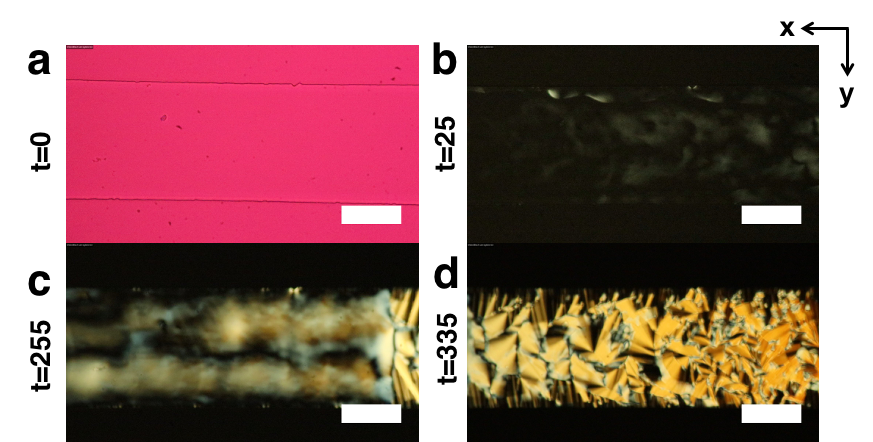}
\caption{\label{fig:12DSCGplasma200-10-TL}\textbf{POM time lapse micrographs capture the isotropic-nematic-M-phase transition of 12~wt.~\% DSCG} within 200$\mu$m wide, 10$\mu$m deep microchannel with degenerate planar anchoring. (\textbf{a}) isotropic phase observed under crossed polarizers with $\lambda$-plate inserted (t=0 minute); (\textbf{b}) crossed polarized micrograph at t=25 minute: faint birefringence is noted, indicating nematic phase; (\textbf{c}) crossed polarized image at t=255 minute captures the appearance of the M-phase (right edge of the micrograph); and (\textbf{d}) at t=355 minute, the channel is covered with the spherulite (M-phase) texture. Scale bar: 100 $\mu$m.}
\end{figure}

\end{document}